\documentclass[12pt]{iopart}
\usepackage{iopams}
\expandafter\let\csname equation*\endcsname\relax
\expandafter\let\csname endequation*\endcsname\relax
\usepackage{amsmath}
\usepackage{amssymb}
\usepackage{graphicx}
\usepackage[T1]{fontenc}
\usepackage{url}
\usepackage{xcolor}
\usepackage{hyperref}
\hypersetup{hidelinks}

\begin{document}

\title{A covariant analysis of massless Dirac fields in LRS spacetimes}

\author{Eduardo Bittencourt${}^{1,2}$\footnote{Corresponding author} and Sante Carloni${}^{3,4,5,6}$ and Giuseppe De Maria${}^{3,4}$ and Peter K. S. Dunsby${}^{2,7,8}$}
\address{${}^{1}$Institute of Physics and Chemistry, Federal University of Itajub\'a, Av.\ BPS 1303, Itajub\'a/MG, Brazil}
\address{${}^{2}$Department of Mathematics and Applied Mathematics, University of Cape Town, Rondebosch 7700, Cape Town, South Africa}
\address{${}^{3}$DIME, Università di Genova, Via all’Opera Pia 15, 16145 Genova, Italy}
\address{${}^{4}$INFN, Sezione di Genova, Via Dodecaneso 33, 16146 Genova, Italy}
\address{${}^{5}$INAF - Osservatorio Astronomico di Roma, via Frascati 33, 00040 - Monte Porzio Catone (Roma), Italy}
\address{${}^{6}$GNFM, Istituto Nazionale di Alta Matematica, P.le Aldo Moro 5, 00185 Roma, Italy}
\address{${}^{7}$Center for Space Research, North-West University, Potchefstroom 2520, South Africa}
\address{${}^{8}$South African Astronomical Observatory, Observatory 7925, Cape Town, South Africa}

\ead{bittencourt@unifei.edu.br, sante.carloni@unige.it, giuseppe.demaria@edu.unige.it, peter.dunsby@uct.ac.za}

\date{\today}

\begin{abstract}
We investigate the massless Dirac field in locally rotationally symmetric (LRS) spacetimes using a covariant spinorial-fluid representation, with the spinor current and spin direction aligned with the LRS frame. When the current has both vorticity and sheet twist, characterized by a constant ratio $\lambda$, the LRS integrability condition relates evolution and propagation derivatives, reducing the field equations to an ordinary differential system that can be integrated exactly. The integrability of the chiral angle forces the heat flux to vanish and the source to saturate the null energy condition along the two radial null directions. A constant boost in the radial Lorentzian two-plane maps the spinor current to a canonical congruence, which is stationary (LRS~I) for $|\lambda|>1$ and irrotational and geodesic (LRS~III) for $|\lambda|<1$. In this frame, the local metric is reconstructed in the Goode--Wainwright--Van den Bergh canonical form. For $|\lambda|>1$ the energy density is positive, and the metric is locally isometric to the horizonless charged Taub--NUT (Brill) metric: in the polar representation, the massless Dirac field supplies a stress--energy tensor algebraically identical to that of the Brill electromagnetic field, with no Maxwell field present. For $|\lambda|<1$, the energy density is negative and the metric is the hyperbolic ($k=-1$) charged NUT metric with a formal negative squared-charge parameter, whose homogeneous slices are of LRS Bianchi type~VIII.
\end{abstract}

\maketitle

\section{Introduction}
\label{sec:introduction}

Spinor fields in curved spacetime provide a natural framework for studying quantum
matter and gravitational geometry simultaneously. Exact self-gravitating Dirac configurations are nevertheless difficult to construct because the Dirac and Einstein equations form a nonlinear coupled system \cite{Finster1999}. Even at the test-field level, reducing the covariant Dirac equation to a tractable problem requires a suitable spin frame, separation of variables, and boundary conditions at horizons and at infinity. This is illustrated by studies on Schwarzschild \cite{BrillWheeler1957,Unruh1976,Cho2003}, Kerr \cite{Chandrasekhar1976,CarterMcLenaghan1979,DolanDempsey2015}, and C-metric \cite{BiniBittencourtGeralico2015} backgrounds.

These difficulties motivate formulations in which the spinorial degrees of freedom are expressed through variables adapted directly to the spacetime geometry. In the polar representation, the Dirac field is described by real covariant quantities and admits an effective imperfect-fluid interpretation. The corresponding energy density, isotropic pressure, energy flux, and anisotropic stress are determined by the polar modulus, derivatives of the chiral angle, and kinematical quantities of the current \cite{Fabbri2021,Fabbri2023,Vignolo2025}. This description is especially useful in treating locally rotationally symmetric (LRS) spacetimes, in which isotropy about a preferred spatial direction reduces all geometrical and matter variables, defined with respect to a preferred frame, to scalar amplitudes. The $(1+1+2)$ formalism writes the field equations in terms of these amplitudes and separates derivatives along the timelike congruence, along the preferred spatial direction, and on the orthogonal two-sheets \cite{Ellis1967,StewartEllis1968,vanElstEllis1996,ClarksonBarrett2003,BetschartClarkson2004,Clarkson2007}.

Here we extend the covariant programme of Ref.~\cite{Vignolo2025} to a massless Dirac field on an LRS spacetime. We specify that this field $\psi$ is not a Weyl or Majorana fermion, but a regular Dirac spinor, for which the scalar bilinear $\bar\psi\psi$ and the pseudoscalar bilinear $i\bar\psi\gamma^5\psi$ do not vanish simultaneously. Under this condition, we can use the polar form of regular spinors described in Ref.~\cite{Fabbri2021}. Weyl and Majorana spinors, which have both bilinears equal to zero, belong instead to the singular classes of the Lounesto classification \cite{Lounesto2001} and are not covered by the present analysis.

Within LRS spacetimes, vorticity and sheet twist cannot both be nonzero in the rest frame of a perfect fluid, so perfect-fluid solutions belong to the classes I--III; imperfect sources also admit configurations in which both are nonzero (the rotating--twisting sector) \cite{Singh2016,Singh2017}. Since the polar Dirac field is generally imperfect, our main question is whether its spinorial constraints are compatible with the rotating--twisting sector and, if so, whether the resulting covariant solution determines an explicit spacetime geometry. We find that, for $m=0$, integrability of the chiral angle selects a single family of rotating--twisting configurations, which we integrate exactly.

We first describe the ``degenerate sectors'' in which either vorticity or sheet twist vanishes, distinguishing the consequences of the zero mass assumption from those specific to simultaneous rotation and twist. We then analyze rotating--twisting geometries covariantly, without imposing a metric ansatz. Compatibility with the spinorial variables closes the characteristic system and permits its exact integration. Because the spinor current has nonzero vorticity, it is not hypersurface orthogonal, and its proper time does not define a time function. Since the ratio $\lambda=\Omega/\xi$ turns out to be constant, a constant boost in the radial Lorentzian two-plane maps the current to a canonical congruence. For $|\lambda|>1$ this congruence is aligned with a timelike Killing vector (LRS~I); for $|\lambda|<1$ it is irrotational and geodesic (LRS~III), so its orthogonal hypersurfaces define a time function on the region covered. In this frame the local metric is reconstructed in the Goode--Wainwright--Van den Bergh canonical form. For $|\lambda|>1$ the reconstructed metric is locally isometric to the horizonless charged Taub--NUT (Brill) metric \cite{Brill1964,Clement2016}. We therefore find that a single massless Dirac field, in the absence of Maxwell fields, is able to mimic the Brill electromagnetic field at the level of the polar stress--energy tensor. For $|\lambda|<1$ the reconstructed metric is the hyperbolic ($k=-1$) member of the charged NUT family with a formal negative squared-charge parameter \cite{PlebanskiDemianski1976}. For nondegenerate $k=+1$ continuations, interior zeros of $v$ correspond to Killing horizons where the polar modulus diverges.

Our classification of the degenerate sectors covers the perfect-spinor case in all three LRS classes and the non-perfect LRS~I and LRS~III cases in the frame defined by the spinor current and spin direction; only the non-perfect LRS~II sector remains outside the present work. As we will see, the restriction to this aligned frame is relevant: when the LRS frame is only coplanar with the spinor current and spin, that is, boosted relative to them in the radial two-plane, non-perfect LRS III configurations exist \cite{VignoloEtAl2026Coplanar}. The boosted LRS III frame, which we will explore in Sec.~\ref{subsec:boosted_frame_interpretation}, is of this type.

The paper is organized as follows. Section~\ref{sec:setup} introduces the covariant variables and the massless spinorial-fluid relations. Section~\ref{sec:branches} examines the degenerate LRS sectors. Section~\ref{sec:generic_branch} derives the characteristic system for the rotating--twisting branch, and Section~\ref{sec:normalized_integration} develops its boosted-frame interpretation, normalized dynamics, and exact integration. Section~\ref{sec:regularity} analyzes regularity and invariant properties, while Section~\ref{sec:metric_map} reconstructs the local metric and gives an elementary representative. Section~\ref{sec:conclusion} summarizes and discusses our results. 

We use natural units $\hbar=c=8\pi G=1$ and metric signature $(+---)$. The connection $\nabla_a$ is the torsion-free Levi--Civita connection and we impose $G_{ab}=T_{ab}$ with vanishing cosmological constant. These assumptions are used explicitly when the Ricci tensor and curvature invariants are reconstructed. Moreover, we use the following convention for the Levi-Civita tensor $\eta_{0123}=\sqrt{-g}$, where $g$ is the determinant of the spacetime metric tensor $g_{ab}$.

\section{Covariant setup and massless spinorial-fluid variables}
\label{sec:setup}

Let $u^a$ be a unit timelike vector and $e^a$ a unit spacelike vector such that
\begin{equation}
        u^a u_a=1, \qquad e^a e_a=-1, \qquad u^a e_a=0.
\end{equation}
The metric on the two-dimensional sheet orthogonal to both congruences is
\begin{equation}
        N_{ab}=h_{ab}+e_a e_b,
\end{equation}
where $h_{ab}=g_{ab}-u_a u_b$ is the three-dimensional projector orthogonal to $u^a$. The sheet alternating tensor is $\varepsilon_{ab}:=\eta_{abcd}u^c e^d$, where $\eta_{abcd}$ is the spacetime volume form. For any scalar $f$ we write $\dot f:=u^a\nabla_a f$, $\hat f:=e^a\nabla_a f$, and $\delta_a f:=N_a{}^b\nabla_b f$. In an LRS spacetime, every physically meaningful sheet vector and projected trace-free sheet tensor vanishes; the nonzero covariant quantities are represented by scalar amplitudes \cite{ClarksonBarrett2003,BetschartClarkson2004,Clarkson2007}. The geometrical scalars used below are
\begin{equation}
        \{\mathcal A,\Theta,\Sigma,\Omega,\phi,\xi,\mathcal E,\mathcal H\},
\end{equation}
where $\mathcal A$ is the radial acceleration, $\Theta$ is the expansion, $\Sigma$ is the shear scalar, $\Omega$ is the vorticity scalar, $\phi$ is the sheet expansion, $\xi$ is the sheet twist, and $\mathcal E,\mathcal H$ are the electric and magnetic Weyl scalars, respectively.

For the matter content, the stress-energy tensor of the effective spinorial fluid is decomposed as
\begin{equation}
        T_{ab} = \mu u_a u_b - p h_{ab} - 2Q u_{(a}e_{b)} +\Pi\left(e_a e_b+\frac{1}{2}N_{ab}\right),
\label{eq:T_decomposition}
\end{equation}
where $\mu$ is the energy density, $p$ is the isotropic pressure, $Q$ is the radial heat-flux scalar, and $\Pi$ is the scalar anisotropic pressure.

In the polar formulation of the Dirac field \cite{Fabbri2021,Vignolo2025}, the spinorial degrees of freedom are represented by a polar modulus $\rho$ and a Yvon--Takabayasi (chiral) angle $\beta$. The modulus is a scalar, whereas $\beta$ is a pseudoscalar under orientation reversal; after an orientation is fixed, both are invariant under proper orthochronous Lorentz-frame changes and, as we will see, may be treated as LRS scalar amplitudes. We assume that the Dirac bilinears inherit the LRS symmetry and align the polar current and spin directions with $u^a$ and $e^a$, respectively. In particular, we will have $\delta_a\rho=\delta_a\beta=0$. This symmetry-inheritance and alignment hypothesis is essential: a merely coplanar geometrical frame permits additional sectors \cite{VignoloEtAl2026Coplanar}. Thus, in the massless limit ($m=0$), the spinorial-fluid variables reduce to
\begin{align}
        \mu &= \frac{\rho}{2}\left(-\frac{\hat\beta}{2}-\Omega\right),\label{eq:mu_massless}\\
        p &= -\frac{\rho}{12}\left(\hat\beta+2\Omega\right),\label{eq:p_massless}\\
        \Pi &= -\frac{\rho}{6}\left(\hat\beta-\Omega\right),\label{eq:Pi_massless}\\
        Q &= -\frac{\rho}{4}\left(\dot\beta+\xi\right).\label{eq:Q_massless}
\end{align}
These equations satisfy $T^{a}{}_{a}=\mu-3p=0$, as expected for a massless Dirac field. Throughout the nontrivial spinor branch, we work on connected open regions where the polar decomposition is regular and $\rho>0$; divisions by $\rho$ below are restricted to such regions.

Under these hypotheses, the spinor polar modulus satisfies
\begin{equation}
        \dot\rho=-\Theta\,\rho, \quad\mbox{and}\quad  \hat\rho=(\mathcal A-\phi)\rho
\label{eq:rho_massless}
\end{equation}
while the derivatives of the chiral angle may be reconstructed as
\begin{equation}
        \hat\beta=\Omega-\frac{6\Pi}{\rho}, \quad\mbox{and}\quad     \dot\beta=-\xi-\frac{4Q}{\rho}.
\label{eq:beta_reconstruction_general}
\end{equation}
Comparison with Ref.~\cite{Vignolo2025} shows that the mass term affects which equations are needed to close the system. When $m\neq0$, $\beta$ enters the field equations algebraically, through $\cos\beta$ and $\sin\beta$, and this additional freedom allows for nontrivial solutions in the LRS I and II classes. For $m=0$, $\beta$ enters only through its derivatives. In the perfect-fluid case, this adds the conditions~\eqref{eq:perfect_beta_conditions} below, and the LRS~I equations become overdetermined (Sec.~\ref{sec:branches}). Without the perfect-fluid condition, the massless LRS~I equations close into a first-order system that admits local solutions (end of Sec.~\ref{sec:branches}); the non-perfect massless LRS~II sector is left for future work. For LRS III, instead, the restriction is stronger: under the assumptions considered below, no nontrivial backreacting massless spinorial source survives, although stealth configurations are not excluded.

\section{Degenerate sectors with \texorpdfstring{$\Omega\xi=0$}{Omega xi = 0}}
\label{sec:branches}

Before studying the rotating-twisting sector, we summarize the features of the covariant equations for LRS I-III spacetimes collected in~\ref{app:lrs_system}. We analyze all three LRS classes under the perfect-fluid condition $Q=\Pi=0$ and then show separately that allowing non-perfect terms does not enlarge the LRS III branch, whereas it does enlarge the LRS~I branch. From now on, we will refer to these cases as \emph{degenerate sectors}.

Under the perfect spinorial fluid assumption, equations~\eqref{eq:Pi_massless} and \eqref{eq:Q_massless} give
\begin{equation}
        \hat\beta=\Omega,
        \qquad
        \dot\beta=-\xi.
\label{eq:perfect_beta_conditions}
\end{equation}
Applying to $\beta$ the LRS scalar relation 
\begin{equation}
        \Omega\,\dot f=\xi\,\hat f.
\label{eq:dot_hat_relation}
\end{equation}
where $f$ represents any LRS scalar, it then forces $\Omega\xi=0$. Thus a perfect massless spinorial fluid cannot belong to the rotating-twisting sector.

In LRS I, with $\xi=0$ and $\Omega\neq0$, the consistency conditions imply
\begin{equation}
\label{LRSI_conditions}
    \Theta=0, \qquad \Sigma=0, \qquad \dot f=0.
\end{equation}
The reduced massless perfect-fluid equations lead to the algebraic condition
\begin{equation}
\label{constr_pf}
        12\mathcal A^2+12\Omega^2+5\rho\Omega=0.
\end{equation}
Propagating this condition along $e^a$ gives
\begin{equation}
        \mathcal A(\mathcal A^2+\Omega^2)=0.
\end{equation}
Since $\Omega\neq0$, one obtains $\mathcal A=0$. However, this is inconsistent with the remaining equations. With $\mathcal A=0$ (hence $\hat{\mathcal A}=0$), the Raychaudhuri equation \eqref{eq:app_Raychaudhuri}, together with \eqref{eq:mu_massless}, \eqref{eq:p_massless}, \eqref{eq:perfect_beta_conditions} and \eqref{LRSI_conditions}, gives
\begin{equation}\label{rho1_LRSI_pf}
\rho=-\tfrac{8}{3}\Omega,
\end{equation}
whereas the equation \eqref{constr_pf}, with $\mathcal A=0$, gives
\begin{equation}\label{rho2_LRSI_pf}
\rho=-\tfrac{12}{5}\Omega.
\end{equation}
Relations \eqref{rho1_LRSI_pf} and \eqref{rho2_LRSI_pf} are verified only if $\rho=\Omega=0$, which contradicts $\Omega\neq0$ and $\rho>0$. Hence the massless perfect-fluid LRS I sector admits no nontrivial regular solution.

In LRS II, one has $\Omega=\xi=0$. Equation~\eqref{eq:perfect_beta_conditions} gives $\dot\beta=\hat\beta=0$, and therefore
\begin{equation}
        \mu=p=Q=\Pi=0.
\end{equation}
Thus, the stress-energy tensor vanishes. The spinor may still be nontrivial at the level of $(\rho,\beta)$, but it does not backreact on the geometry. We call the configuration \emph{stealth} when the polar variables remain nonzero but $T_{ab}=0$; the term \emph{vacuum} is reserved for the geometry (or for the trivial spinor). The LRS II massless perfect branch is therefore a stealth or vacuum sector.

In LRS III, the condition is $\Omega=0$ and $\xi\neq0$. The LRS identities then imply
\begin{equation}
        \hat f=0,
        \qquad
        \mathcal A=0,
        \qquad
        \phi=0.
\end{equation}
For a massless perfect spinor, equation~\eqref{eq:perfect_beta_conditions} gives $\hat\beta=0$ and $\dot\beta=-\xi$. Again, the stress-energy tensor vanishes, so the LRS III perfect branch is a stealth or vacuum configuration. Moreover, even if the spinorial fluid is allowed to be non-perfect in LRS III, the condition $\hat f=0$ implies $\hat\beta=0$, hence $\Pi=0$ from equation~\eqref{eq:Pi_massless}. The LRS matter constraint then forces $Q=0$. Thus the non-perfect LRS III sector collapses to the perfect one under the aligned spinorial-frame assumption used in this section. This conclusion does not apply to a geometrical LRS III frame boosted relative to the spinor-current frame \cite{VignoloEtAl2026Coplanar}.

In LRS~I, by contrast, the non-perfect sector is, in general, not empty ( see also \cite{Vignolo2025}). With $\xi=0$, $\Omega\neq0$ and $\dot f=0$, the matter constraint~\eqref{eq:app_matter_constraint} gives $Q\Omega^2=0$, hence $Q=0$, and equation~\eqref{eq:Q_massless} is then consistent with $\dot\beta=0$. Writing $b:=\hat\beta$, equations~\eqref{eq:mu_massless}--\eqref{eq:Pi_massless} give
\begin{equation}
        \mu=-\frac{\rho}{4}\left(b+2\Omega\right), \qquad p=\frac{\mu}{3}, \qquad \Pi=-\frac{\rho}{6}\left(b-\Omega\right),
\label{eq:LRSI_nonperfect_matter}
\end{equation}
and the evolution equations~\eqref{eq:app_dot_phi}, \eqref{eq:app_dot_xi}, \eqref{eq:app_dot_Sigma} and the constraint~\eqref{eq:app_H_constraint} fix the Weyl scalars algebraically,
\begin{equation}
        \mathcal E=-\mathcal A\phi+2\Omega^2+\frac{\rho}{4}\left(b+\Omega\right), \qquad \mathcal H=\Omega\left(2\mathcal A+\phi\right).
\label{eq:LRSI_nonperfect_weyl}
\end{equation}
The Raychaudhuri equation~\eqref{eq:app_Raychaudhuri}, the propagation equations~\eqref{eq:app_hat_Omega} and \eqref{eq:app_hat_phi}, the density equation~\eqref{eq:rho_massless} and the momentum conservation law~\eqref{eq:app_momentum_conservation} then give the closed first-order system
\begin{align}
        \hat{\mathcal A}&=\mathcal A^2-\mathcal A\phi+2\Omega^2+\frac{\rho}{4}\left(2\Omega+b\right), &
        \hat\phi&=-\frac{\phi^2}{2}-\mathcal A\phi+2\Omega^2+\frac{\rho}{2}\left(\Omega+b\right),\nonumber\\
        \hat\Omega&=-\Omega\left(\mathcal A+\phi\right), &
        \hat\rho&=\left(\mathcal A-\phi\right)\rho, \qquad
        \hat b=\mathcal A\left(b+2\Omega\right)+\Omega\phi.
\label{eq:LRSI_nonperfect_system}
\end{align}
After these substitutions, the remaining equations, in particular the propagation equations~\eqref{eq:app_hat_E} and \eqref{eq:app_hat_H} for the Weyl scalars, are satisfied identically. Hence equations~\eqref{eq:LRSI_nonperfect_system} are a free first-order system for $(\mathcal A,\phi,\Omega,\rho,b)$, and by the standard existence theorem for ordinary differential equations local solutions exist for arbitrary initial data with $\Omega\neq0$ and $\rho>0$. The energy density is positive when $b<-2\Omega$. The perfect sub-case $b=\Omega$ is not invariant under this flow: requiring $\hat b=\hat\Omega$ gives $\phi=-2\mathcal A$, and propagating this relation reproduces the constraint~\eqref{constr_pf} and therefore the contradiction derived above. A global analysis of system~\eqref{eq:LRSI_nonperfect_system} and the reconstruction of the corresponding stationary metrics are left for future work.

Hence, among the sectors considered here, only the non-perfect LRS~I sector and the rotating--twisting sector can carry a massless spinorial source with $T_{ab}\neq0$. The latter is analyzed in the rest of the paper.

\section{Compatibility with the rotating-twisting branch}
\label{sec:generic_branch}

In the rotating-twisting sector, where $\Omega\xi\neq0$, the compatibility condition \eqref{eq:dot_hat_relation} can be written as
\begin{equation}
        \hat f=\lambda\dot f,
        \quad\mbox{with}\quad
        \lambda:=\frac{\Omega}{\xi}.
\label{eq:hat_lambda_dot}
\end{equation}
The geometrical constraint and the massless density equation imply\footnote{The equations used in this short derivation are displayed in~\ref{app:lrs_system}.}
\begin{equation}
        \phi=\lambda\left(\Sigma+\frac{2}{3}\Theta\right),
        \quad\mbox{and}\quad
        \mathcal A=\lambda\left(\Sigma-\frac{1}{3}\Theta\right).
\label{eq:generic_geometric_relations_setup}
\end{equation}
Using these relations in the evolution equations for $\Omega$ and $\xi$ gives
\begin{equation}
        \frac{\dot\Omega}{\Omega} = \frac{\dot\xi}{\xi} = -\left(2\Sigma+\frac{1}{3}\Theta\right).
\label{eq:lambda_constancy_derivation}
\end{equation}
All statements in this section are understood on a connected component of the rotating--twisting region, so $\Omega\xi\neq0$ throughout. Consequently,
\begin{equation}
        \dot\lambda=0,
        \qquad
        \hat\lambda=\lambda\dot\lambda=0.
\end{equation}
Thus, the dot and hat derivatives of every LRS scalar in the rotating-twisting sector are related by a constant on each such component. Equivalently, equation \eqref{eq:dot_hat_relation} becomes
\begin{equation}
        (e^a-\lambda u^a)\nabla_a f=0,
\label{eq:characteristic_condition}
\end{equation}
so every LRS scalar is constant along $e^a-\lambda u^a$ and depends on a single variable. As in Ref.~\cite{VanDenBergh2017}, this reduces the scalar field equations to ordinary differential equations.

The reduction introduced by $\lambda$ is consistent with the scalar commutator. With our sign conventions, one obtains
\begin{equation}
        \widehat{\dot f}-\dot{\hat f} = \mathcal A\dot f -  \left(\Sigma-\frac{1}{3}\Theta\right)\hat f .
\label{eq:LRS_commutator_scalar}
\end{equation}
Together with equations~\eqref{eq:hat_lambda_dot} and \eqref{eq:generic_geometric_relations_setup}, it yields only the derived relation~\eqref{eq:characteristic_wave_equation}, which is consistent with equation~\eqref{eq:hat_lambda_dot}.

\subsection{Algebraic reconstruction}

We introduce the auxiliary combinations
\begin{equation}
        X:=\Sigma+\frac{2}{3}\Theta, \qquad Y:=\Sigma-\frac{1}{3}\Theta,
\label{eq:XY_def}
\end{equation}
so that
\begin{equation}
        \Theta=X-Y, \qquad \Sigma=\frac{1}{3}(X+2Y).
\label{eq:XY_inverse}
\end{equation}
The compatibility between the spinorial density equation and the LRS constraint gives
\begin{equation}
        \phi=\lambda X, \qquad   \mathcal A=\lambda Y,
\label{eq:basic_reconstruction}
\end{equation}
in agreement with equation~\eqref{eq:generic_geometric_relations_setup}.

We define $\Delta:=3\Pi-\lambda\rho\xi$, which measures the violation of the LRS relation~\eqref{eq:dot_hat_relation} by the chiral angle, as shown below. In terms of $\Delta$, the formulae \eqref{eq:app_massless_mu}--\eqref{eq:app_massless_density} take the form
\begin{align}
        \mu&=\frac{1}{2}\Delta-\frac{1}{4}\lambda\rho\xi,
\label{eq:mu_Delta}\\
        p&=\frac{1}{6}\Delta-\frac{1}{12}\lambda\rho\xi,
\label{eq:p_Delta}\\
        \Pi&=\frac{1}{3}\left(\Delta+\lambda\rho\xi\right),
\label{eq:Pi_Delta}\\
        Q&=\frac{\lambda}{1+\lambda^2}\Delta,
\label{eq:Q_Delta}\\
        \mathcal H&=(\lambda^2-1)\xi(X+2Y),
\label{eq:H_Delta}\\
        \mathcal E&=(1-\lambda^2) \left[ XY-2\xi^2+\frac{\Delta}{2(1+\lambda^2)} \right].
\label{eq:E_Delta}
\end{align}
The chiral angle is reconstructed from
\begin{equation}
        \hat\beta = -\lambda\xi-\frac{2\Delta}{\rho}, \qquad  \dot\beta = -\xi-\frac{4\lambda}{(1+\lambda^2)\rho}\Delta.
\label{eq:beta_Delta}
\end{equation}
These relations do not by themselves make $\beta$ an LRS scalar: that property is the symmetry-inheritance hypothesis of Sec.~\ref{sec:setup}, and its consequence~\eqref{eq:beta_LRS_constraint} is imposed in the next subsection. Once the algebraic relations are used, every LRS scalar is a function of $(\rho,\xi,X,Y,\Delta)$; the differential equations of Subsection~\ref{subsec:closed_red_dyn} then evolve $(\rho,\xi,X,Y)$.

\subsection{Chiral-angle compatibility}

Since $\beta$ is an LRS scalar, it must also satisfy
\begin{equation}
        \hat\beta=\lambda\dot\beta.
\label{eq:beta_LRS_constraint}
\end{equation}
Combining equations~\eqref{eq:beta_Delta} and \eqref{eq:beta_LRS_constraint} gives
\begin{equation}
        \hat\beta-\lambda\dot\beta = \frac{2(\lambda^2-1)}{1+\lambda^2} \frac{\Delta}{\rho}.
\end{equation}
For a regular spinor density, $\rho\neq0$, one obtains
\begin{equation}
        (\lambda^2-1)\Delta=0.
\label{eq:branch_constraint}
\end{equation}
Thus either $\lambda^2=1$ or $\Delta=0$\footnote{This dichotomy is a consequence of symmetry inheritance by $\beta$ and need not hold in an unaligned or non-symmetry-inheriting spin frame.}.

A regular rotating-twisting spinor cannot realize the first possibility. Indeed, the evolution equation for $\phi$ gives
\begin{equation}
        \dot X = -\frac{1}{2}X(X+2Y) +2\xi^2 -\frac{\Delta}{1+\lambda^2},
\label{eq:Xdot_general}
\end{equation}
whereas the Raychaudhuri-type mixed equation gives
\begin{equation}
        \dot X+(\lambda^2-1)\dot Y = -\frac{1}{2}X^2 -\lambda^2XY  +(\lambda^2-1)Y^2 +2\lambda^2\xi^2 -\frac{1}{2}\Delta +\frac{1}{4}\lambda\rho\xi .
\label{eq:Ray_XY_Delta}
\end{equation}
For $\lambda^2=1$, these two equations differ by the term $\lambda\rho\xi/4$. Since the regular generic branch has $\lambda\xi\rho\neq0$, this is impossible. Thus, the case $\lambda^2=1$ can only survive as a zero-density boundary limit, which we discard.

Therefore, the regular branch is\footnote{Here and below, ``regular branch'' denotes~\eqref{eq:regular_Delta_zero_branch} on regions with $\rho>0$. The stronger requirement that an orbit be of fixed sign, which excludes zeros of $\xi/\rho$, is introduced in Subsection~\ref{subsec:twist_normalized_phase_space}.}
\begin{equation}
        \Delta=0, \qquad \lambda^2\neq1.
\label{eq:regular_Delta_zero_branch}
\end{equation}
Under this assumption, the matter variables reduce to
\begin{equation}
        \mu=-\frac{1}{4}\lambda\rho\xi, \qquad p=\frac{1}{3}\mu,  \qquad \Pi=\frac{\lambda}{3}\rho\xi, \qquad Q=0.
\label{eq:matter_Delta_zero}
\end{equation}
Hence, the spinorial source saturates the null energy condition along the two preferred radial null directions,
\begin{equation}
        \mu+p+\Pi=0.
\label{eq:matter_constraints_Delta_zero}
\end{equation}
This radial saturation should not be confused with the full null energy condition, which additionally requires $\mu\geq0$, as shown explicitly in Sec.~\ref{subsec:ricci_invariant_distinction}.

The relations above imply that the LRS matter consistency condition
\begin{equation}
        (\mu+p+\Pi)\xi\Omega=Q(\Omega^2+\xi^2)
\end{equation}
is identically satisfied, which also explains why this case can retain simultaneous rotation and spatial twist without the nonzero heat flux required in the strict energy-condition sector of the general imperfect-fluid analysis of References~\cite{Singh2016,Singh2017}. The anisotropic pressure reconstructed from equation \eqref{eq:matter_Delta_zero} also evolves in agreement with the conservation equation. Thus, the constraint $\Delta=0$ is dynamically preserved.

In this branch, the chiral angle satisfies
\begin{equation}
        \dot\beta=-\xi, \qquad \hat\beta=-\lambda\xi,
\label{eq:beta_Delta_zero}
\end{equation}
so equation~\eqref{eq:beta_LRS_constraint} is identically fulfilled. For completeness, we conclude this section by also giving an expression of the Weyl scalars in our variables:
\begin{equation}
        \mathcal H=(\lambda^2-1)\xi(X+2Y), \qquad \mathcal E=(1-\lambda^2)(XY-2\xi^2).
\label{eq:weyl_Delta_zero}
\end{equation}

\subsection{Closed reduced dynamics}
\label{subsec:closed_red_dyn}
The closed reduced system follows directly from the $\xi$ evolution, the spinor density equation, the $\phi$ evolution equation, and the Raychaudhuri-type mixed equation. In terms of our free variables, it is given by
\begin{align}
        \dot X &= -\frac{X^2}{2}-XY+2\xi^2,
\label{eq:final_X}\\
        \dot Y &= -XY+Y^2+2\xi^2 + \frac{\lambda\rho\xi}{4(\lambda^2-1)},
\label{eq:final_Y}\\
        \dot\rho &= -(X-Y)\rho,
\label{eq:final_rho}\\
        \dot\xi &= -(X+Y)\xi.
\label{eq:final_xi}
\end{align}
In~\ref{app:lrs_system} we report the explicit derivation of the above system. Substitution of equations~\eqref{eq:basic_reconstruction}, \eqref{eq:matter_Delta_zero}, \eqref{eq:weyl_Delta_zero}, and \eqref{eq:final_X}--\eqref{eq:final_xi} into the remaining LRS equations gives identities, so the branch \eqref{eq:regular_Delta_zero_branch} generates no further algebraic condition. Hence, the equations above completely determine the evolution of the rotating--twisting sector.

Note that this consistent branch has no regular equilibrium point. If $\dot\rho=\dot\xi=0$ with $\rho\xi\neq0$, then equations~\eqref{eq:final_rho} and \eqref{eq:final_xi} imply $X=Y=0$. But then equation~\eqref{eq:final_X} gives $\dot X=2\xi^2\neq0$. Thus, the chiral-angle compatibility condition removes the equilibrium line of the unconstrained system.

The evolution parameter of the system (\ref{eq:final_X})--(\ref{eq:final_xi}) is the proper time $\tau$ of the spinor current. Because the current has vorticity, it is not hypersurface orthogonal, so $\tau$ does not define a time function, and the system does not by itself describe the evolution of spatial hypersurfaces. Moreover, by equation ~\eqref{eq:characteristic_condition}, every scalar also varies along $e^a$. Section~\ref{sec:normalized_integration} resolves this with a constant boost to a congruence that is stationary for $|\lambda|>1$ and irrotational for $|\lambda|<1$; we analyze the boosted equations before integrating them.

\section{Boosted-frame dynamics and exact integration}
\label{sec:normalized_integration}

We now apply a Lorentz boost in the $(u^a,e^a)$ plane, i.e.\ a hyperbolic rotation of $u^a$ and $e^a$ among themselves. A boost of this kind leaves the sheets invariant, so it maps the $1+1+2$ split, and the LRS scalar amplitudes, to those of another adapted frame.

A change of frame does not change the causal structure of the spacetime; in particular, it cannot remove closed timelike curves, which are present on the full angular domain of the $|\lambda|>1$ realization (Sec.~\ref{subsec:elementary_metric_representative}). The boost instead supplies a congruence adapted to the geometry. For $|\lambda|<1$ it is irrotational and geodesic, so its orthogonal hypersurfaces are the level sets of a time function on the region covered. The problem of whether they are actual Cauchy surfaces \cite{BernalSanchez2003,BernalSanchez2005} will not be established here. For $|\lambda|>1$, the congruence is aligned with a timelike Killing vector. Two features are specific to the present solution: the rapidity is constant, because $\lambda$ is constant by~\eqref{eq:lambda_constancy_derivation}, whereas for a generic imperfect fluid in the rotating--twisting sector the rapidity between the fluid frame and the canonical frame need not be constant \cite{VanDenBergh2017}; and the stress tensor of the compatible branch is invariant under the boost, as shown in equation~\eqref{eq:boost_invariant_stress_tensor} below.

We analyze the reduced system in two complementary ways. The qualitative analysis determines its invariant sets, their stability, and the asymptotic limits of its orbits. The exact integration gives the covariant variables as explicit functions of the characteristic parameter; these functions are used in the regularity analysis of Sec.~\ref{sec:regularity} and in the metric reconstruction of Sec.~\ref{sec:metric_map}.

\subsection{Boosted-frame representation}
\label{subsec:boosted_frame_interpretation}

Consider a constant boost in the preferred Lorentzian two-plane, mapping the $(u^a,e^a)$ basis into a new one $(\widetilde{u}^a,\widetilde{e}^a)$ such that
\begin{equation}
\begin{split}
\widetilde u^{a} &=\cosh\eta\,u^{a}+\sinh\eta\,e^{a},\\
\widetilde e^{a} &=\sinh\eta\,u^{a}+\cosh\eta\,e^{a}.
\end{split}
\label{boost}
\end{equation}
For any LRS scalar $f$, the derivatives in the boosted frame are
\begin{align}
D_{\widetilde u}f &=\cosh\eta\left(\dot f+\tanh\eta\,\hat f\right), & D_{\widetilde e}f &=\cosh\eta\left(\hat f+\tanh\eta\,\dot f\right).
\label{eq:boosted_scalar_derivatives}
\end{align}

We first consider $|\lambda|>1$. Choosing $\tanh\eta=-1/\lambda$ gives
\begin{equation}
D_{\widetilde u}f=0, \qquad \widetilde\xi=\widetilde\Theta=\widetilde\Sigma=0,
\label{eq:boost_lrsi_vanishing}
\end{equation}
whereas the nonzero two-plane quantities transform as
\begin{equation}
(\widetilde\Omega,\widetilde\phi,\widetilde{\mathcal A})=\frac{\sqrt{\lambda^2-1}}{|\lambda|} (\Omega,\phi,\mathcal A).
\label{eq:boost_lrsi_remaining}
\end{equation}
The boosted congruence is therefore LRS~I. All scalar equations become spatial propagation equations along $\widetilde e^a$, while the geometry is stationary with respect to $\widetilde u^a$. Since $\widetilde\Omega\neq0$, the congruence $\widetilde u^a$ is not hypersurface orthogonal. Only this range of $\lambda$ contains fixed-sign orbits with $\mu>0$ (see subsection~\ref{subsec:twist_normalized_phase_space}).

For $|\lambda|<1$, the choice $\tanh\eta=-\lambda$ instead yields
\begin{equation}
D_{\widetilde e}f=0, \qquad \widetilde\Omega = \widetilde{\mathcal A} = \widetilde\phi=0,
\label{eq:boost_lrsiii_vanishing}
\end{equation}
and
\begin{equation}
(\widetilde\Theta,\widetilde\Sigma,\widetilde\xi,\widetilde X,\widetilde Y) =\sqrt{1-\lambda^2}\,(\Theta,\Sigma,\xi,X,Y).
\label{potentials}
\end{equation}
This selects an LRS III congruence for which the scalar equations are purely evolutionary. Since $\widetilde\Omega=\widetilde{\mathcal A}=0$, the congruence $\widetilde u^a$ is geodesic and hypersurface orthogonal, and its orthogonal hypersurfaces are the level sets of a time function. Since $\rho$ and $\beta$ are Lorentz scalars, $\widetilde\rho=\rho$, $\widetilde\beta=\beta$, and $D_{\widetilde e}\rho=D_{\widetilde e}\beta=0$. The regular fixed-sign solutions in this range have $\mu<0$, whatever the sign of $\xi$, as shown in Subsection~\ref{subsec:twist_normalized_phase_space}. The limiting case $\lambda^2=1$ is excluded: either construction would require an infinite rapidity, consistent with the obstruction expressed by equation~\eqref{eq:branch_constraint}.

The matter variables are unchanged by either boost, but only because the stress tensor on the compatible branch has the special form
\begin{equation}
T_{ab}=\mu\left(u_a u_b-e_a e_b-N_{ab}\right) =\mu\left(\widetilde u_a\widetilde u_b -\widetilde e_a\widetilde e_b-N_{ab}\right).
\label{eq:boost_invariant_stress_tensor}
\end{equation}
Consequently,
\begin{equation}
\widetilde\mu=\mu, \qquad \widetilde p=p, \qquad \widetilde\Pi=\Pi, \qquad \widetilde Q=0.
\end{equation}
This boost invariance is not a generic property of imperfect fluids, but it holds for the effective fermion fluid\footnote{The boosted congruence is not the spinor current, and the polar-fluid identifications~\eqref{eq:mu_massless}--\eqref{eq:rho_massless} hold only in the original bilinear frame \cite{VignoloEtAl2026Coplanar}. Equation~\eqref{eq:boost_invariant_stress_tensor} expresses the stress tensor of the same Dirac field in the boosted frame. The boost changes only the frame used to describe the solution; it does not act on the spinor field. Statements about the metric, its curvature invariants, the energy conditions and the causal structure are therefore frame independent, whereas the polar identifications (the meaning of $\rho$, $\beta$ and of the kinematical quantities of the current) refer to the original bilinear frame $(u^a,e^a)$ \cite{VignoloEtAl2026Coplanar}.}.

The boosted LRS III description also permits a direct comparison with the coplanar-frame formulation of Ref.~\cite{VignoloEtAl2026Coplanar,esposito2026diracfieldslrsiii}. For $|\lambda|<1$, let $\eta_{\mathrm{cp}}$ denote the rapidity between its geometrical frame and the spinor-current frame. The boost above gives $\tanh\eta_{\mathrm{cp}}=-\lambda$, and hence $\sinh\eta_{\mathrm{cp}} =-\lambda/\sqrt{1-\lambda^2}$ and $\widetilde{\xi}=\sqrt{1-\lambda^2}\,\xi$. Setting $m=0$ in its massless LRS III matter relations therefore gives
\[
 \mu=\frac{\rho\widetilde{\xi}\sinh\eta_{\mathrm{cp}}}{4}
    =-\frac{\lambda\rho\xi}{4}, \qquad p=\frac{\mu}{3}, \qquad \Pi=-\frac{4\mu}{3}, \qquad Q=0,
\]
in agreement with the source obtained here. Our result supplies an exactly integrated massless member of that broader coplanar system, together with its metric reconstruction and invariant characterization.

\subsection{Twist-normalized qualitative dynamics}
\label{subsec:twist_normalized_phase_space}

Let $\tau$ be local proper time along $u^a$ (a local parameter on each integral curve, not necessarily a global time coordinate) and define the auxiliary time variable $s$ and its corresponding derivative as\footnote{The parameter $s$ requires $\xi\neq0$. By equation~\eqref{eq:final_xi}, $\xi=0$ is an invariant set, so $\xi$ cannot vanish at finite $\tau$ while $X+Y$ stays bounded; on fixed-sign orbits, $\xi\neq0$ for all $\tau\in\mathbb R$ by equation~\eqref{eq:xi_solution_final}.}
\begin{equation}
\frac{ds}{d\tau}=\xi, \qquad f':=\frac{df}{ds}=\frac{1}{\xi}\frac{df}{d\tau}.
\label{eq:normalized_parameter_definition}
\end{equation}
Now, we introduce a set of variables normalized by the sheet twist, which is nonzero throughout the rotating--twisting branch, as follows
\begin{equation}
x:=\frac{X}{\xi}, \qquad y:=\frac{Y}{\xi}, \qquad r:=\frac{\rho}{\xi}.
\label{eq:normalized_variables_final}
\end{equation}
The inverse relations are $X=x\xi$, $Y=y\xi$, and $\rho=r\xi$. From equation \eqref{eq:final_xi}, the dimensional scale satisfies
\begin{equation}
\frac{\xi'}{\xi}=-(x+y), 
\label{eq:scale_equation_final}
\end{equation}
whereas the scale-free variables, by using equations \eqref{eq:final_X}--\eqref{eq:final_rho}, obey the triangular autonomous system
\begin{align}
x'&=2+\frac{x^2}{2},\label{eq:xprime_final}\\
y'&=2+2y^2+2ar,\label{eq:yprime_final}\\
r'&=2yr.\label{eq:uprime_final}
\end{align}
where we define $a:=\lambda/[8(\lambda^2-1)]$. Since $\lambda\neq0$ and $\lambda^2\neq1$, one also has $a\neq0$. Thus the scale-free flow separates from the reconstruction of $\xi$. Moreover, $x'>0$ everywhere, so the full three-dimensional system has no finite equilibrium point, in agreement with the direct argument following equations~\eqref{eq:final_X}--\eqref{eq:final_xi}.

The transformation $\lambda\mapsto-1/\lambda$ leaves $a$ invariant and exchanges the LRS I and LRS III ranges. Hence each twist-normalized phase portrait has a dual realization in the two boost regimes, although their energy signs and geometrical interpretations differ.

\subsubsection{Finite reduced dynamics.} The autonomous subsystem in $(y,r)$ has the critical point
\begin{equation}
P_*:\qquad (y_*,r_*)=\left(0,-\frac{1}{a}\right).
\label{eq:reduced_center}
\end{equation}
Its Jacobian is
\begin{equation}
J(P_*)= \begin{pmatrix}
0&2a\\[2pt]
-2/a&0
\end{pmatrix},
\qquad \operatorname{spec}J(P_*)=\{+2i,-2i\}.
\label{eq:reduced_center_jacobian}
\end{equation}
As the first integral below confirms, $P_*$ is a nonlinear center. It is not an equilibrium spacetime because in the full state space it lifts to the invariant orbit
\begin{equation}
\mathcal L_*:\qquad y=0, \qquad r=-\frac{1}{a}, \qquad x'=2+\frac{x^2}{2}>0.
\label{eq:invariant_reference_orbit}
\end{equation}

The transverse dynamics becomes elementary upon setting
\begin{equation}
v:=\frac{1}{r}, \quad \mbox{and}\quad \pi_v:=-\frac{2y}{r}.
\label{eq:v_p_definition}
\end{equation}
Then, equations~\eqref{eq:yprime_final} and \eqref{eq:uprime_final} reduce to
\begin{equation}
v'=\pi_v, \qquad \pi_v'=-4v-4a,
\label{eq:shifted_oscillator_system}
\end{equation}
or, equivalently,
\begin{equation}
v''+4v=-4a.
\label{eq:v_linear_equation_final}
\end{equation}
The oscillator is centered at $(v,\pi_v)=(-a,0)$ and has conserved energy
\begin{equation}
\pi_v^2+4\left(v+a\right)^2=4B^2.
\label{eq:oscillator_energy}
\end{equation}
Equivalently,
\begin{equation}
\mathcal I:=\frac{y^2+2ar+1}{r^2} =B^2-a^2, \qquad y^2=\mathcal I r^2-2ar-1.
\label{eq:yu_first_integral_final}
\end{equation}
The orbit classes are therefore
\begin{align}
\mathcal I<0 &\quad\Longleftrightarrow\quad B^2<a^2, &&v\ \hbox{has fixed sign},\nonumber\\
\mathcal I=0 &\quad\Longleftrightarrow\quad B^2=a^2, &&v\ \hbox{is tangent to zero},\nonumber\\
\mathcal I>0 &\quad\Longleftrightarrow\quad B^2>a^2, &&v\ \hbox{crosses zero}.
\label{eq:orbit_classification}
\end{align}
Only the fixed-sign class remains in one regular twist-normalized chart over a complete oscillator period. The surface $v=0$ is the boundary $|r|=\infty$, not an additional finite state of equations~\eqref{eq:xprime_final}--\eqref{eq:uprime_final}. At a generic zero of $v$, both $y$ and $r$ diverge, and the solution leaves the regular-spinor sector.

\begin{figure}[htbp]
\centering
\includegraphics[width=\linewidth]{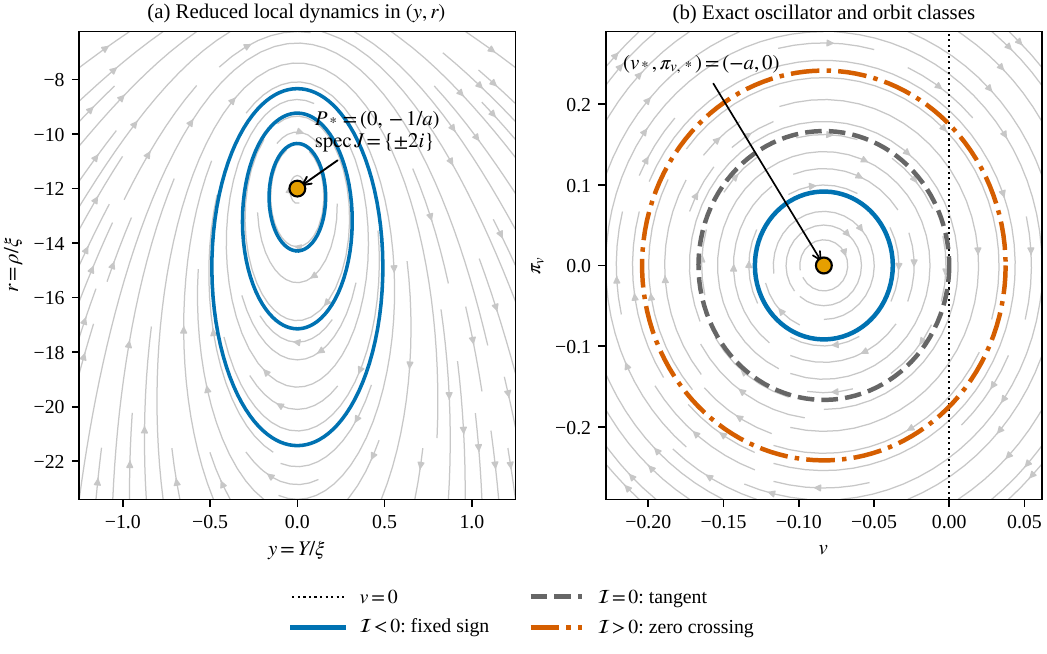}
\caption{Local qualitative analysis of equations~\eqref{eq:xprime_final}--\eqref{eq:uprime_final} for $a=1/12$, corresponding to $\lambda=2$ (LRS I) or $\lambda=-1/2$ (LRS III). Panel~(a) shows the reduced $(y,r)$ flow and the center $P_*=(0,-12)$. Panel~(b) shows the equivalent oscillator portrait in $(v,\pi_v)$. The solid blue curves denote the fixed-sign class $\mathcal I<0$, the dashed grey curve the tangent class $\mathcal I=0$, and the dash-dotted vermilion curve the zero-crossing class $\mathcal I>0$. The dotted vertical line is $v=0$; arrows indicate increasing $s$.}
\label{fig:xi_normalized_local}
\end{figure}

Figure~\ref{fig:xi_normalized_local} illustrates the reduced flow for $a=1/12$. This value is realized by $\lambda=2$ in the LRS I range and, under the duality above, by $\lambda=-1/2$ in the LRS III range. Panel~(a) displays the center $P_*=(0,-12)$ and its closed level sets. These curves are not periodic orbits of the full system because $x$ evolves monotonically. Panel~(b) shows the same solutions in the regular oscillator variables: the inner blue ellipse has $\mathcal I<0$, the dashed ellipse is tangent to $v=0$, and the outer vermilion ellipse crosses it twice.

The fixed-sign class ($\mathcal I<0$) also determines the physical sign of the source. Since its oscillator does not cross the origin, we have $\operatorname{sgn}v=-\operatorname{sgn}a$. Because $\rho>0$ and $v=\xi/\rho$, equations~\eqref{eq:matter_Delta_zero} then give
\begin{equation}
\operatorname{sgn}\xi=\operatorname{sgn}v, \qquad \operatorname{sgn}\mu =\operatorname{sgn}(\lambda^2-1).
\label{eq:energy_sign_fixed_sign_branch}
\end{equation}
Thus the globally regular LRS I realization has positive energy density, whereas its LRS III dual has negative energy density. For $a=1/12$, one has $v<0$ and therefore $\xi<0$. This sign argument is not a linearization: the transformation~\eqref{eq:v_p_definition} maps the nonlinear system~\eqref{eq:yprime_final}--\eqref{eq:uprime_final} exactly onto the linear oscillator~\eqref{eq:v_linear_equation_final}, whose general solution is given in~\eqref{eq:v_solution_final}. The classification~\eqref{eq:orbit_classification} therefore holds globally, not only near $P_*$. 

\subsubsection{Critical sets at infinity}

For the global analysis \cite{WainwrightEllis1997,Matsue2018}, let $\boldsymbol q=(x,y,r)$ and compactify $\mathbb R^3$ to the closed unit ball using the coordinates:
\begin{equation}
\boldsymbol z =\frac{\boldsymbol q}{\sqrt{1+\|\boldsymbol q\|^2}}, \qquad \varpi=\sqrt{1-\|\boldsymbol z\|^2}.
\label{eq:xi_poincare_ball_map}
\end{equation}
The homogeneous parts of the vector field are
\begin{align}
\boldsymbol F_0&=(2,2,0),\nonumber\\
\boldsymbol F_1&=(0,2ar,0),\nonumber\\
\boldsymbol F_2&=\left(\frac{x^2}{2},2y^2,2yr\right).
\label{eq:xi_homogeneous_parts}
\end{align}
Since the system is quadratic, define the regularized parameter by $d\eta_{\rm c}/ds=\varpi^{-1}$. The compactified equations are
\begin{equation}
\frac{d\boldsymbol z}{d\eta_{\rm c}} =\sum_{j=0}^{2}\varpi^{2-j} \left[ \boldsymbol F_j(\boldsymbol z) -\boldsymbol z\bigl(\boldsymbol z\mathbin{\cdot} \boldsymbol F_j(\boldsymbol z)\bigr)\right].
\label{eq:xi_compactified_system}
\end{equation}
At $\varpi=0$, only $\boldsymbol F_2$ survives. The boundary-equilibrium condition
\begin{equation}
\boldsymbol F_2(\boldsymbol z) -\boldsymbol z\bigl(\boldsymbol z\mathbin{\cdot} \boldsymbol F_2(\boldsymbol z)\bigr)=\boldsymbol0, \qquad \|\boldsymbol z\|=1,
\label{eq:xi_boundary_equilibrium_condition}
\end{equation}
gives the following critical sets
\begin{align}
M_\pm&=(\pm1,0,0),\nonumber\\
\mathcal C_-&:\quad z_x=0, \qquad z_y^2+z_r^2=1,\nonumber\\
\mathcal C_+&:\quad z_y=\frac{z_x}{4}, \qquad \frac{17}{16}z_x^2+z_r^2=1.
\label{eq:xi_boundary_equilibrium_sets}
\end{align}
The two circles meet at
\begin{equation}
D_\pm=(0,0,\pm1).
\label{eq:xi_degenerate_boundary_points}
\end{equation}

To classify these sets, define
\begin{equation}
h(\boldsymbol z) :=\boldsymbol z\mathbin{\cdot}\boldsymbol F_2(\boldsymbol z) =\frac{1}{2}z_x^3+2z_y^3+2z_yz_r^2.
\label{eq:xi_radial_h}
\end{equation}
Near the boundary,
\begin{equation}
\frac{d\varpi}{d\eta_{\rm c}} =-h(\boldsymbol z)\varpi+O(\varpi^2).
\label{eq:xi_radial_equation}
\end{equation}
Combining this radial rate with the angular linearization gives Table~\ref{tab:xi_boundary_equilibria}. The zero rate on either circle is tangent to that circle.

\begin{table}[htbp]
\centering
\caption{Critical sets at infinity of the $\xi$-normalized system. Rates are listed in the radial direction and the two angular directions for increasing $\eta_{\rm c}$, equivalently increasing $s$. The final column gives the temporal interpretation in the boosted LRS III frame; in the LRS I frame the same sets describe spatial propagation limits. An $s$-source ($s$-sink) is a set approached by nearby orbits at the lower (upper) end of their range of $s$, i.e.\ as $\eta_{\rm c}\to-\infty$ ($+\infty$).}
\label{tab:xi_boundary_equilibria}
\small
\begin{tabular}{llll}
\hline
Set & Linear rates & Type & LRS III limit\\
\hline
$M_-$ & $(1/2,1/2,1/2)$ & $s$-source & flat Milne type\\
$M_+$ & $(-1/2,-1/2,-1/2)$ & $s$-sink & flat Milne type\\
$\mathcal C_-$ & $(-2z_y,-2z_y;0)$ & attracting/repelling & flat LRS Kasner/Taub\\
$\mathcal C_+$ & $(-z_x/2,z_x/2;0)$ & saddle & non-flat LRS Kasner\\
$D_\pm$ & $(0,0,0)$ & degenerate & no regular limit\\
\hline
\end{tabular}
\end{table}

On $\mathcal C_-$, both normal rates are negative for $z_y>0$ and positive for $z_y<0$, so the two semicircles are normally attracting and repelling, respectively. On $\mathcal C_+$ the nonzero normal rates have opposite signs, making the circle saddle-like away from $D_\pm$. All leading linear rates vanish at $D_\pm$. Within the exact orbit family, these directions arise when $v$ has a double zero and $r=1/v$ diverges; they do not belong to the regular fixed-sign sector. A directional blow-up would be required only for a complete resolution of this nonregular boundary flow.

The asymptotic geometry can be read in a clear way from the twist-normalized projective directions (see~\ref{app:critical_point_reconstruction} for details). In particular, for $|\lambda|<1$, let $\gamma=\sqrt{1-\lambda^2}$. Equations~\eqref{eq:XY_inverse} and
\eqref{potentials} give
\begin{equation}
\widetilde\Theta=\gamma\xi(x-y), \qquad \widetilde\Sigma=\frac{\gamma\xi}{3}(x+2y).
\label{eq:projective_theta_sigma_direct}
\end{equation}
Writing $\boldsymbol q=\mathcal{R}\boldsymbol z+o(\mathcal{R})$ as $\mathcal{R}\to\infty$, one obtains
\begin{equation}
\frac{\widetilde\Sigma}{\widetilde\Theta} \longrightarrow \frac{z_x+2z_y}{3(z_x-z_y)}  =
\begin{cases}
\phantom{-}1/3, & M_\pm,\\
-2/3, & \mathcal C_-\setminus\{D_\pm\},\\
\phantom{-}2/3, & \mathcal C_+\setminus\{D_\pm\}.
\end{cases}
\label{eq:projective_shear_ratio_direct}
\end{equation}
At the same nondegenerate sets,
\begin{equation}
\frac{\widetilde\xi}{\widetilde\Theta}=O(\mathcal{R}^{-1}), \qquad \frac{\mu}{\widetilde\Theta^2}=O(\mathcal{R}^{-1}),
\label{eq:projective_source_decay_direct}
\end{equation}
and the normalized pressure and anisotropic stress vanish at the same rate, since $p=\mu/3$ and $\Pi=-4\mu/3$. Thus, the limiting directional dynamics is twist-free and source-free even when $z_r\neq0$. The polar modulus, encoded in $z_r$, therefore enters only at subleading order and does not affect the leading-order geometry: the asymptotic regimes are vacuum regimes, even along directions with $z_r\neq0$.

The directional expansion rates then give the power-law triples (see~\ref{app:critical_point_reconstruction})
\begin{equation} 
M_\pm:(0,1,1), \qquad \mathcal C_-:(1,0,0), \qquad \mathcal C_+:\left(-\frac13,\frac23,\frac23\right).
\label{eq:projective_power_law_summary}
\end{equation}
They correspond, respectively, to flat Milne-type, flat LRS Kasner/Taub, and non-flat LRS Kasner limits in the boosted LRS III frame. At $D_\pm$ one has $z_x-z_y=0$, so this reconstruction is undefined, consistent with the nonregular character of those points. None of the nondegenerate critical sets has vanishing asymptotic shear ratio; hence there is no isotropic FLRW fixed point. In particular, the exponents $(0,1,1)$ of the sets $M_\pm$ describe flat spacetime foliated by hypersurfaces $\mathbb R\times H^2$, namely, the tensor product of two-dimensional Milne space with a line, with $\widetilde\Sigma/\widetilde\Theta=1/3\neq0$. The isotropic Milne universe would have exponents $(1,1,1)$.

\begin{figure}[htbp]
\centering
\includegraphics[width=\linewidth]{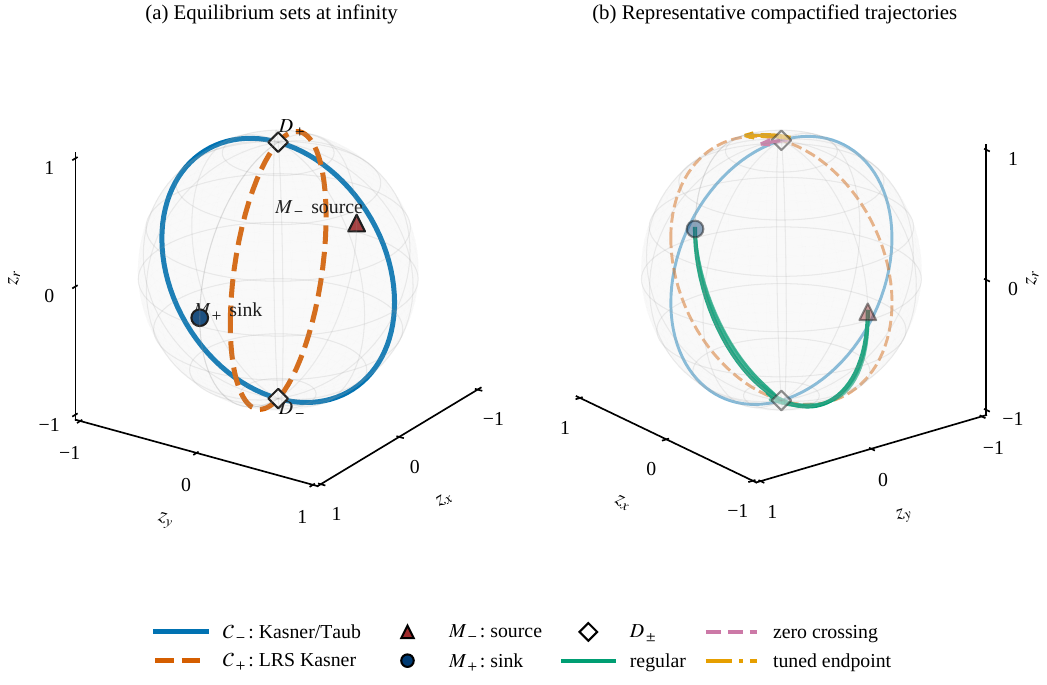}
\caption{Global Poincar\'e-ball representation of the $\xi$-normalized system for $a=1/12$, corresponding to $\lambda=2$ (LRS I) or $\lambda=-1/2$ (LRS III). Panel~(a) shows the boundary sets: $\mathcal C_-$ is the solid blue circle, $\mathcal C_+$ the dashed vermilion circle, $M_-$ the triangular source, $M_+$ the circular sink, and $D_\pm$ the white diamonds. Panel~(b) shows representative trajectories: solid green curves are regular fixed-sign orbits joining $M_-$ to $M_+$, the dashed purple curve has zero-crossing endpoints on $\mathcal C_-$, and the dash-dotted orange curve has a tuned endpoint on $\mathcal C_+$. Regular orbits run from $M_-$ to $M_+$ as $s$ increases.}
\label{fig:xi_normalized_global}
\end{figure}

Figure~\ref{fig:xi_normalized_global} displays the boundary sets and representative orbits of the compactified system for $a=1/12$. The green fixed-sign trajectories have bounded $(y,r)$ and satisfy $M_-\longrightarrow M_+$, for increasing $s$. Along them, $|x|\to\infty$ while $y$ and $r$ remain finite. After restoring the scale, all connection and curvature scalars decay at both ends. In the positive-energy LRS I realization ($\lambda=2$), these are the two spatial ends of the stationary geometry; in the LRS III dual ($\lambda=-1/2$), they are the two orientations of the same flat Milne-type temporal limit. A simple zero of $v$ produces a nonregular trajectory toward $\mathcal C_-$, whereas a simultaneous zero of $v$ and pole of $x$ selects a direction on $\mathcal C_+$. A double zero approaches $D_\pm$. These are projective statements and do not make a zero of $v$ part of the regular spinor branch.

The same compactified equations and stability rates therefore apply in both boost regimes. Only their interpretation differs: for $|\lambda|>1$ the parameter $s$ labels position along the spin direction of a stationary geometry, whereas for $|\lambda|<1$ it labels time, and the limits of Table~\ref{tab:xi_boundary_equilibria} apply. The twist normalization thus treats both ranges of $\lambda$ with a single autonomous system.

\subsection{Exact integration}
\label{subsec:twist_normalized_exact_solution}

In the twist-normalized variables, the system integrates in closed form. Equation~\eqref{eq:xprime_final} gives
\begin{equation}
        x(s)=2\tan(s-s_x),
\label{eq:x_solution_final}
\end{equation}
where $s_x$ is an integration constant. The shifted oscillator
\eqref{eq:v_linear_equation_final} gives
\begin{equation}
        v(s) =B\cos\left[2(s-s_u)\right] -a,
\label{eq:v_solution_final}
\end{equation}
and hence
\begin{equation}
        r(s) =\frac{1}{B\cos\left[2(s-s_u)\right] -a},
\label{eq:u_solution_final}
\end{equation}
\begin{equation}
        y(s) =\frac{B\sin\left[2(s-s_u)\right]}{B\cos\left[2(s-s_u)\right] -a}.
\label{eq:y_solution_final}
\end{equation}

The twist $\xi$, which sets the scale of all dimensional variables, follows from equation~\eqref{eq:scale_equation_final}. Since $x=2\tan(s-s_x)$ and $y=-v'/(2v)$,
\begin{equation}
        \frac{\xi'}{\xi} = -2\tan(s-s_x) + \frac{v'}{2v}.
\end{equation}
Thus, on an interval where $v$ has definite sign, one has
\begin{equation}
        \xi(s) =\xi_0 \cos^2(s-s_x)\sqrt{|v(s)|}.
\label{eq:xi_solution_final}
\end{equation}
Because $\rho>0$ and $v=\xi/\rho$, the integration constant must satisfy $\operatorname{sgn}(\xi_0)=\operatorname{sgn}(v)$ on every fixed-sign interval. Consequently, the polar modulus is
\begin{equation}
        \rho(s) = r(s)\xi(s) = \frac{|\xi_0|\cos^2(s-s_x)}{\sqrt{|v(s)|}}.
\label{eq:rho_solution_final}
\end{equation}
The original variables are then
\begin{equation}
        X(s)=2\xi(s)\tan(s-s_x),
\label{eq:X_solution_final}
\end{equation}
\begin{equation}
        Y(s) = \xi(s)\frac{B\sin\left[2(s-s_u)\right]}{B\cos\left[2(s-s_u)\right] -a}.
\label{eq:Y_solution_final}
\end{equation}
Finally, $\Theta$ and $\Sigma$ can be reconstructed from equation~\eqref{eq:XY_inverse}.

The chiral angle also integrates directly. Since $\dot\beta=-\xi$, one has
\begin{equation}
        \beta'=-1, \qquad \beta(s)=\beta_0-s.
\label{eq:beta_solution_final}
\end{equation}
The remaining covariant and matter variables are reconstructed algebraically from equations~\eqref{eq:basic_reconstruction}, \eqref{eq:matter_Delta_zero}, and \eqref{eq:weyl_Delta_zero}.

The relation between the normalized parameter $s$ and the original parameter $\tau$ is obtained from equation~\eqref{eq:normalized_parameter_definition}. Therefore, using equation~\eqref{eq:xi_solution_final}, one obtains
\begin{equation}
        \tau-\tau_0= \frac{1}{\xi_0} \int^s\frac{d\bar s}{        \cos^2(\bar s-s_x) \sqrt{\left|B\cos\left[2(\bar s-s_u) \right]-a\right|}}.
\label{eq:tau_quadrature_explicit_final}
\end{equation}
Thus, the regular rotating-twisting massless spinorial branch is integrable in parametric form. Direct differentiation of equations~\eqref{eq:x_solution_final}--\eqref{eq:Y_solution_final} reproduces equations~\eqref{eq:scale_equation_final}--\eqref{eq:uprime_final}; this also fixes the relative signs in $y=-v'/(2v)$ and in the reconstruction of $\rho$.

\subsection{Global relation between the normalized and physical time parameters}
\label{subsec:global_time_map}

Because $s$ ranges over a finite interval, it is not evident that the solutions of equation~\eqref{eq:tau_quadrature_explicit_final} are defined for all values of the proper time $\tau$ of the current. This subsection shows that, on fixed-sign orbits, the finite $s$-interval corresponds to $\tau\in\mathbb R$, so the solutions do not end at finite proper time. To make this explicit, let us define
\begin{equation}
        \theta_x:=s-s_x, \quad\mbox{and} \quad \psi:=s-s_u,
\label{eq:theta_psi_v_time_map}
\end{equation}
and consider the fundamental interval
\begin{equation}
        \mathcal I_x := \left(s_x-\frac{\pi}{2},s_x+\frac{\pi}{2}\right).
\label{eq:fundamental_s_interval}
\end{equation}
On this interval, the function $v$ has fixed sign and is bounded away from zero if and only if
\begin{equation}
        |B|<|a|.
\label{eq:v_fixed_sign_global_condition}
\end{equation}
On such a branch, one has $\operatorname{sgn}v = \operatorname{sgn}\xi_0 = -\operatorname{sgn}a$. Hence, the map $s\mapsto\tau$ is increasing when $\xi_0>0$ and decreasing when $\xi_0<0$.

At the two endpoints of $\mathcal I_x$, the function $v$ approaches the same finite value
\begin{equation}
        v_{\rm end} := v\left(s_x\pm\frac{\pi}{2}\right) = -a-B\cos\left[2(s_x-s_u)\right].
\label{eq:v_endpoint_value}
\end{equation}
By defining
\begin{equation}
        \delta:=\frac{\pi}{2}-|s-s_x|,
\end{equation}
one obtains
\begin{equation}
        \xi(s) = \xi_0\sqrt{|v_{\rm end}|}\,\delta^2 \left[1+\mathcal{O}(\delta)\right].
\label{eq:xi_endpoint_asymptotic}
\end{equation}
Equation~\eqref{eq:tau_quadrature_explicit_final} then gives, as $\delta\to0$,
\begin{equation}
        |\tau| = \frac{1}{|\xi_0|\sqrt{|v_{\rm end}|}\,\delta} \left[1+\mathrm{o}(1)\right],
\label{eq:tau_endpoint_asymptotic}
\end{equation}
up to an additive shift of $\tau$. Hence,
\begin{equation}
        s\rightarrow s_x\pm\frac{\pi}{2} \quad \Longleftrightarrow \quad \tau \rightarrow \pm\operatorname{sgn}(\xi_0)\,\infty .
\label{eq:s_tau_global_bijection}
\end{equation}
Thus, although $s$ takes values in a finite interval, it parametrizes the complete range $\tau\in\mathbb R$, with its orientation determined by the sign of $\xi_0$. In particular,
\begin{equation}
        \frac{\pi}{2}-|s-s_x| = \frac{1}{|\xi_0|\sqrt{|v_{\rm end}|}\,|\tau|} \left[1+\mathrm{o}(1)\right].
\label{eq:s_inverse_endpoint_asymptotic}
\end{equation}

The phase-aligned elliptic quadrature, its inverse, and the exceptional parameter cases are collected in~\ref{app:elliptic_quadrature}.

\section{Covariant regularity and invariant characterization}
\label{sec:regularity}

This section examines the regularity and the invariant characterization of the integrated solutions. Subsection~\ref{subsec:covariant_regularity} bounds the connection scalars and the curvature invariants on fixed-sign orbits; Subsection~\ref{subsec:ricci_invariant_distinction} shows that the Ricci tensor distinguishes the solution invariantly from the rotating--twisting imperfect fluids of Refs.~\cite{Singh2016,Singh2017}.

\subsection{Regularity of the integrated branch}
\label{subsec:covariant_regularity}

We restrict first to a connected interval on which
\begin{equation}
        |v(s)|\geq v_{\min}>0.
\label{eq:v_regular_lower_bound}
\end{equation}
This condition excludes zeros of the normalization variable $v=1/r=\xi/\rho$ and ensures that the spinorial density and the adapted LRS connection remain bounded. Using equations~\eqref{eq:xi_solution_final} and \eqref{eq:rho_solution_final}, one finds the particularly simple identity
\begin{equation}
        \rho(s)\xi(s) = \xi_0|\xi_0|\cos^4\theta_x.
\label{eq:rho_xi_exact_regular}
\end{equation}
Since the source~\eqref{eq:matter_Delta_zero} depends on $\rho$ and $\xi$ only through the product $\rho\xi$, it is independent of $v(s)$:
\begin{equation}
        \mu(s) = -\frac{\lambda\xi_0|\xi_0|}{4}\cos^4\theta_x, \qquad p=\frac{\mu}{3}, \qquad \Pi=-\frac{4\mu}{3}, \qquad Q=0.
\label{eq:matter_exact_regular}
\end{equation}
All matter components are bounded on the closure of the fundamental interval and vanish at its two infinite-time endpoints.

The kinematical variables can be written as
\begin{align}
        \xi &= \xi_0\cos^2\theta_x\sqrt{|v|},\label{eq:xi_regular_explicit}\\
        \rho &= \frac{|\xi_0|\cos^2\theta_x}{\sqrt{|v|}},\label{eq:rho_regular_explicit}\\
        X &= \xi_0\sin(2\theta_x)\sqrt{|v|},\label{eq:X_regular_explicit}\\
        Y &= \frac{|\xi_0|B\cos^2\theta_x\sin(2\psi)}{\sqrt{|v|}},\label{eq:Y_regular_explicit}
\end{align}
where $\theta_x$ and $\psi$ were defined in equation~\eqref{eq:theta_psi_v_time_map}. Equations~\eqref{eq:v_regular_lower_bound}--\eqref{eq:Y_regular_explicit} show that $X,Y,\rho$ and $\xi$ are bounded. The reconstruction formulae \eqref{eq:XY_inverse} and \eqref{eq:basic_reconstruction} then imply that $\Theta$, $\Sigma$, $\mathcal A$, $\phi$, and $\Omega$ are bounded as well. Thus, none of the Ricci rotation coefficients of the adapted LRS frame diverges on the regular branch.

The trace of the stress-energy tensor vanishes because $\mu=3p$. Einstein's equations, therefore, give
\begin{equation}
        R=0, \qquad R_{ab}=T_{ab}.
\label{eq:R_zero_Ricci_equals_T}
\end{equation}
Introducing the metric in the preferred Lorentzian two-plane,
\begin{equation}
        P_{ab}:=u_a u_b-e_a e_b,
\label{eq:preferred_plane_metric}
\end{equation}
Equation~\eqref{eq:matter_exact_regular} yields the compact Ricci-tensor form
\begin{equation}
        R_{ab} = \mu\left(P_{ab}-N_{ab}\right).
\label{eq:Ricci_exact_spinor_branch}
\end{equation}
Consequently,
\begin{equation}
        R_{ab}R^{ab} = 4\mu^2 = \frac{\lambda^2\xi_0^4}{4}        \cos^8\theta_x,
\label{eq:Ricci_squared_exact_regular}
\end{equation}
which is finite everywhere and tends to zero at both ends of the physical-time axis.

The Weyl scalars can also be written without any denominator involving $v$. Direct substitution into equation~\eqref{eq:weyl_Delta_zero} gives
\begin{align}
    \mathcal E &= 2(1-\lambda^2)\xi_0|\xi_0|\cos^3\theta_x \left[B\sin\theta_x\sin(2\psi) -v\cos\theta_x \right],\label{eq:E_exact_regular_s}\\
    \mathcal H &= 2(\lambda^2-1)\xi_0|\xi_0|\cos^3\theta_x \left[v\sin\theta_x +B\cos\theta_x\sin(2\psi) \right].\label{eq:H_exact_regular_s}
\end{align}
These quantities are bounded for all finite $s$.  With the LRS normalization used in this work,
\begin{equation}
        C_{abcd}C^{abcd} = 12\left(\mathcal E^2-\mathcal H^2\right),
\label{eq:Weyl_squared_LRS}
\end{equation}
and the Kretschmann scalar is
\begin{equation}
        R_{abcd}R^{abcd} = 12\left(\mathcal E^2-\mathcal H^2\right) +2R_{ab}R^{ab}.
\label{eq:Kretschmann_LRS_regular}
\end{equation}
Equations~\eqref{eq:Ricci_squared_exact_regular}, \eqref{eq:E_exact_regular_s}, and \eqref{eq:H_exact_regular_s} show that the orthonormal-frame components of the Ricci and Weyl tensors are bounded on the branch~\eqref{eq:v_regular_lower_bound}; hence every zeroth-order scalar polynomial formed algebraically from the Riemann tensor is bounded there. This statement alone does not establish extendibility or geodesic completeness.

On every globally fixed-sign branch, $v_{\rm end}\neq0$, and the endpoint behavior is independent of the sign of $v_{\rm end}$. Combining equations~\eqref{eq:s_inverse_endpoint_asymptotic} and \eqref{eq:xi_regular_explicit}--\eqref{eq:H_exact_regular_s}, one obtains
\begin{align}
        X&=O(|\tau|^{-1}), & \xi,\rho,Y&=O(|\tau|^{-2}),\label{eq:kinematic_tau_asymptotics}\\
        \mu,p,\Pi&=O(|\tau|^{-4}), & \mathcal E,\mathcal H&=O(|\tau|^{-3}), \label{eq:curvature_tau_asymptotics}\\
        R_{ab}R^{ab}&=O(|\tau|^{-8}), & R_{abcd}R^{abcd}&=O(|\tau|^{-6}). \label{eq:invariant_tau_asymptotics}
\end{align}
The curvature therefore approaches zero at both asymptotic ends. Moreover, on the branch \eqref{eq:v_regular_lower_bound}, every covariant scalar is smooth in $s$, while
\begin{equation}
        \dot f=\xi\frac{df}{ds}, \qquad \hat f=\lambda\xi\frac{df}{ds}.
\label{eq:dot_hat_s_regularity}
\end{equation}
For every fixed differential order, repeated dot and hat derivatives consequently remain bounded. This supplies a covariant regularity check stronger than testing only the displayed lowest-order polynomial invariants, while still falling short of a global extension theorem.

The endpoint-degenerate aligned case $\chi_{\rm e}=0$, defined in~\ref{app:elliptic_quadrature}, also has bounded curvature. There $v=2B\cos^2s$ with $B\neq0$, and the characteristic parameter behaves as $|\tau|\propto\delta^{-2}$. The source combination \eqref{eq:rho_xi_exact_regular} remains valid, and all curvature scalars stay finite. The case $\alpha_{\rm e}=0$ is not resolved by the metric reconstruction. At the internal zero of $v$, the Ricci and Weyl scalars remain bounded, but $\rho$ and the adapted connection scalar $Y$ diverge. Moreover, equation~\eqref{eq:metric_scale_factor_solution} gives $F\propto\sqrt{|v|}$, so the canonical coframe displayed in Sec.~\ref{sec:metric_map} degenerates at that surface. Whether this is removable by a different coordinate system and spin frame requires a separate extension analysis. This case is therefore not included in the regular-spinor sector.

Finally, bounded curvature and the global range $\tau\in\mathbb R$ establish that the preferred characteristic contains no finite-parameter scalar curvature singularity. In the B2/LRS III representation, $w=(\tau-\tau_0)/\Gamma_\lambda+w_0$ is a natural time coordinate and covers $\mathbb R$ on the regular branch. In the B1/LRS I representation, by contrast, $w$ is spatial, and the canonical fiber direction is timelike (its tangent vector is $\partial_z$). Neither the local metric reconstruction nor the unbounded characteristic range alone proves geodesic completeness, the existence of Cauchy surfaces, or global hyperbolicity. These properties require specified coordinate ranges and topology together with a separate analysis of all causal geodesics.

\subsection{Invariant distinction from the strict-NEC branch}
\label{subsec:ricci_invariant_distinction}

The massless spinorial solution belongs to the broad kinematical sector $\Omega\xi\neq0$, but it is invariantly distinct from the strict-energy-condition solutions discussed in Refs.~\cite{Singh2016,Singh2017}. This distinction can be formulated directly in terms of the Ricci tensor and therefore does not depend on the choice of coordinates or on a boost of the adapted LRS frame. On any open non-conformally-flat region where the LRS isotropy singles out the preferred $2+2$ splitting, the associated radial null directions are therefore geometric up to interchange and reciprocal rescaling.

Introduce the two preferred radial null vectors
\begin{equation}
        k_+^a:=u^a+e^a, \qquad k_-^a:=u^a-e^a, \qquad k_+^a k_{-a}=2.
\label{eq:preferred_radial_null_pair}
\end{equation}
For the general LRS stress-energy tensor \eqref{eq:T_decomposition}, Einstein's equations imply
\begin{equation}
        R_{ab}k_\pm^a k_\pm^b = T_{ab}k_\pm^a k_\pm^b = \mu+p+\Pi\pm2Q.
\label{eq:Ricci_radial_null_general}
\end{equation}
The trace term in Einstein's equations does not contribute because $k_\pm^a$ is null. In the present branch, equations~\eqref{eq:matter_constraints_Delta_zero} and \eqref{eq:matter_Delta_zero} give
\begin{equation}
        R_{ab}k_+^a k_+^b = R_{ab}k_-^a k_-^b =0.
\label{eq:Ricci_radial_null_saturated}
\end{equation}
Thus the null energy condition is saturated along both preferred radial null directions.

This marginal behavior is transparent from the exact Ricci tensor \eqref{eq:Ricci_exact_spinor_branch}. Its mixed form has two double eigenvalues,
\begin{equation}
        \operatorname{spec}\!\left(R^a{}_b\right) = \{\mu,\mu,-\mu,-\mu\},
\label{eq:Ricci_eigenvalues_spinor}
\end{equation}
where the $+\mu$ eigenspace is the preferred Lorentzian two-plane spanned by $u^a$ and $e^a$, while the $-\mu$ eigenspace is the LRS sheet. The restriction of $R_{ab}$ to the preferred two-plane is proportional to its two-dimensional metric. It therefore annihilates every null vector contained in that plane.

More generally, let $n^a$ be any null vector and decompose it into its preferred two-plane and sheet parts. Equation~\eqref{eq:Ricci_exact_spinor_branch} gives
\begin{equation}
        R_{ab}n^a n^b = -2\mu N_{ab}n^a n^b.
\label{eq:Ricci_general_null_spinor}
\end{equation}
Since $N_{ab}n^a n^b\leq0$, the NEC holds whenever $\mu\geq0$, and equality occurs precisely when the null vector has no sheet component. The source therefore satisfies the NEC marginally in the radial directions and strictly in nonradial null directions when $\mu>0$.

A useful boost-invariant discriminator is the product
\begin{equation}
        \mathfrak D_R := \left(R_{ab}k_+^a k_+^b\right)  \left(R_{cd}k_-^c k_-^d\right) = (\mu+p+\Pi)^2-4Q^2.
\label{eq:Ricci_null_discriminant}
\end{equation}
Under a boost in the $(u^a,e^a)$ plane, $k_+^a$ and $k_-^a$ are rescaled by opposite factors, so their product in equation~\eqref{eq:Ricci_null_discriminant} is unchanged. The spinorial branch satisfies
\begin{equation}
        \mathfrak D_R=0,
\label{eq:Ricci_discriminant_spinor_zero}
\end{equation}
whereas the strict radial NEC assumed in the previous rotating-twisting imperfect-fluid sector implies
\begin{equation}
        R_{ab}k_+^a k_+^b>0,  \qquad  R_{ab}k_-^a k_-^b>0, \qquad    \mathfrak D_R>0.
\label{eq:Ricci_discriminant_previous_positive}
\end{equation}
No coordinate transformation, tetrad rotation, or finite boost can map a region with $\mathfrak D_R=0$ to one with $\mathfrak D_R>0$. Therefore, the present solution is not locally isometric to any member of that strict-NEC branch.

This distinction is independent of the exceptional kinematical case $\lambda^2=1$, which is excluded by equation~\eqref{eq:regular_Delta_zero_branch}. The vanishing Ricci discriminant follows instead from equations~\eqref{eq:matter_Delta_zero} and~\eqref{eq:matter_constraints_Delta_zero}. The spinorial solution therefore belongs to a Ricci-degenerate boundary branch of the rotating--twisting LRS sector and is not locally isometric to any member of the strict-NEC family. This invariant argument does not exclude a limiting relation in which $\mathfrak D_R\to0$. The reconstruction in Sec.~\ref{sec:metric_map} places the solution in the Goode--Wainwright--Van den Bergh B2 class for $|\lambda|<1$ and B1 class for $|\lambda|>1$, with $k<0$ and $k>0$, respectively. This canonical classification is compatible with the Ricci distinction established above.

\section{Explicit metric reconstruction}
\label{sec:metric_map}

The exact covariant solution determines not only the connection and curvature scalars but also a local coordinate representative of the spacetime. To make this explicit, recall that the derivatives of the preferred frame vectors are
\begin{align}
\nabla_a u_b &= -\mathcal A u_a e_b +Y e_a e_b +\frac{1}{2}X N_{ab} +\Omega\varepsilon_{ab},\label{eq:nabla_u_map}\\
\nabla_a e_b &= -\mathcal A u_a u_b +Y e_a u_b +\frac{1}{2}\phi N_{ab} +\xi\varepsilon_{ab}.\label{eq:nabla_e_map}
\end{align}
Thus $\{\mathcal A,X,Y,\Omega,\phi,\xi\}$ provide the local connection data of an adapted LRS tetrad. The boost of Sec.~\ref{sec:normalized_integration} places these data directly in one of the canonical metric classes of Goode--Wainwright and Van den Bergh \cite{GoodeWainwright1986,VanDenBergh2017}.

\subsection{Canonical form and metric functions}
\label{subsec:canonical_metric_form}

Define
\begin{equation}
        \epsilon_\lambda:=\operatorname{sgn}(\lambda^2-1), \qquad    \Gamma_\lambda:=\sqrt{|1-\lambda^2|}.
\label{eq:metric_branch_sign_scale}
\end{equation}
The regular branch has $\lambda^2\neq1$, so $\epsilon_\lambda=\pm1$ and $\Gamma_\lambda>0$. Introduce a complex coordinate $\zeta$ on a constant-curvature two-space and define
\begin{align}
        e^{-\Psi_k} &:= 1+\frac{k}{4}\zeta\bar\zeta, & d\Sigma_k^2    &:= e^{2\Psi_k}d\zeta\,d\bar\zeta,
\label{eq:canonical_sheet_metric}\\
        \omega_k &:= e^{\Psi_k}\operatorname{Im}  \left(\zeta\,d\bar\zeta\right), & \vartheta &:= dz+q\omega_k .
\label{eq:canonical_fibre_form}
\end{align}
Here $k$ is the Gaussian curvature parameter of $d\Sigma_k^2$ and $q\neq0$. With the discrete Goode--Wainwright parameter written as $\epsilon_\lambda$, the canonical line element is
\begin{equation}
        d\ell^2 = -\epsilon_\lambda\,dw^2 +\epsilon_\lambda F^2(w)\vartheta^2 -H^2(w)d\Sigma_k^2 .
\label{eq:canonical_reconstructed_metric}
\end{equation}
For $\epsilon_\lambda=-1$, this is class B2: $w$ is timelike and the canonical congruence is LRS III. For $\epsilon_\lambda=+1$, it is class B1: $w$ is spacelike, the fibre direction is timelike, and the canonical congruence is LRS I. The null class B3 would correspond to the excluded limit
$\lambda^2=1$.

Choose the orientation of $w$ so that
\begin{equation}
        \frac{ds}{dw}=\Gamma_\lambda\xi .
\label{eq:metric_s_w_relation}
\end{equation}
In the LRS III sector, the longitudinal and transverse expansion rates give $F_{,w}/F=-\widetilde Y$ and $H_{,w}/H=\widetilde X/2$. In the LRS I sector, the radial orientation associated with the chosen coordinate $w$ is $e_{\mathrm{can}}^{a} =\operatorname{sgn}(\lambda)\widetilde e^{a}=\partial_w$. Consequently, $\mathcal A_{\mathrm{can}} =\operatorname{sgn}(\lambda)\widetilde{\mathcal A} =-F_{,w}/F$ and $\phi_{\mathrm{can}} =\operatorname{sgn} (\lambda) \widetilde\phi =2H_{,w}/H$. Using the boost relations, both sectors therefore reduce to
\begin{equation}
        \frac{d\ln F}{ds}=-y, \qquad \frac{d\ln H}{ds}=\frac{x}{2}.
\label{eq:metric_scale_factor_equations}
\end{equation}
Since $y=-v'/(2v)$ and $x=2\tan(s-s_x)$, integration gives
\begin{equation}
        F(s)=F_0\sqrt{|v(s)|}, \qquad H(s)=H_0\sec(s-s_x),
\label{eq:metric_scale_factor_solution}
\end{equation}
where $F_0,H_0>0$. Moreover,
\begin{equation}
        \frac{qF(s)}{H^2(s)} = \Gamma_\lambda\xi(s), \qquad  frac{qF_0}{H_0^2}=\Gamma_\lambda\xi_0,
\label{eq:metric_twist_normalization}
\end{equation}
where the sign of $q$ changes if the sheet orientation is reversed. In the B2 frame, $qF/H^2$ equals $\widetilde\xi$. In the B1 canonical radial orientation specified above, it equals $\Omega_{\mathrm{can}}=\operatorname{sgn}(\lambda)\widetilde\Omega$. The orientation factor is relevant when $\lambda<0$.

Thus the general fixed-sign representative can be displayed directly as
\begin{equation}
\begin{split}
        d\ell^2={}& -\epsilon_\lambda\,dw^2 +\epsilon_\lambda F_0^2  \left|-a +B\cos\left[2(s(w)-s_u)\right]\right|     \left[ dz+q e^{\Psi_k}  \operatorname{Im} \left(\zeta\,d\bar\zeta\right) \right]^2\\
        &-H_0^2\sec^2\left[s(w)-s_x\right] e^{2\Psi_k}d\zeta\,d\bar\zeta .
\end{split}
\label{eq:general_metric_explicit_parametric}
\end{equation}
The constants $q$, $H_0$, and $k$ are related by equations~\eqref{eq:metric_twist_normalization} and \eqref{eq:metric_k_constant} below.

Finally, equation~\eqref{eq:metric_s_w_relation} and equation~\eqref{eq:xi_solution_final} give
\begin{equation}
        w-w_0 = \frac{1}{\Gamma_\lambda\xi_0} \int^s \frac{d\bar s} {\cos^2(\bar s-s_x) \sqrt{\left|-a +B \cos\left[2(\bar s-s_u)\right]\right|}}.
\label{eq:metric_w_quadrature}
\end{equation}
Thus $w=(\tau-\tau_0)/\Gamma_\lambda+w_0$ along the characteristic parametrization, up to the chosen orientation. The inverse elliptic map in~\ref{app:elliptic_quadrature} consequently also provides $s=s(w)$ in the phase-aligned subfamily. Equations \eqref{eq:canonical_reconstructed_metric}, \eqref{eq:general_metric_explicit_parametric}, and \eqref{eq:metric_w_quadrature} constitute an explicit parametric reconstruction of the full fixed-sign solution.

\subsection{Metric integrability and the curvature parameter}
\label{subsec:metric_integrability}

It remains to verify that the curvature parameter $k$ is constant and that no additional restriction is generated by the Cartan equations. In the canonical orthonormal frame, equation~\eqref{eq:boost_invariant_stress_tensor} requires
\begin{equation}
        G_{\hat0\hat0}=\mu,  \qquad G_{\hat1\hat1}=-\mu, \qquad     G_{\hat2\hat2}=G_{\hat3\hat3}=\mu, \qquad G_{\hat0\hat1}=0.
\label{eq:metric_Einstein_components_required}
\end{equation}
Using $\rho=r\xi$ and equation~\eqref{eq:normalized_variables_final}, the density can be written as
\begin{equation}
        \mu = -2\epsilon_\lambda ar (\Gamma_\lambda\xi)^2.
\label{eq:metric_mu_normalized}
\end{equation}
The $\hat0\hat0$ component of the Einstein tensor of equation~\eqref{eq:canonical_reconstructed_metric} is
\begin{equation}
        \frac{G_{\hat0\hat0}}{(\Gamma_\lambda\xi)^2} = \epsilon_\lambda \left(1+xy-\frac{x^2}{4}\right) +\frac{k}{H^2(\Gamma_\lambda\xi)^2}.
\label{eq:metric_Einstein_00}
\end{equation}
Equating equations~\eqref{eq:metric_mu_normalized} and \eqref{eq:metric_Einstein_00} fixes
\begin{equation}
        \frac{k}{H^2}  = -\epsilon_\lambda(\Gamma_\lambda\xi)^2     \left(1+xy-\frac{x^2}{4}+2ar\right).
\label{eq:metric_k_constraint}
\end{equation}
This is an integrability condition, not a new evolution equation. Indeed, let $\epsilon_v:=\operatorname{sgn}v$ on a fixed-sign interval. Direct substitution of equations~\eqref{eq:x_solution_final}--\eqref{eq:y_solution_final} gives the identity
\begin{equation}
        \cos^2(s-s_x)|v(s)| \left(1+xy-\frac{x^2}{4}+2ar\right)= \epsilon_v \left\{ a + B \cos\left[2(s_x-s_u)\right] \right\}.
\label{eq:metric_k_constant_identity}
\end{equation}
Consequently,
\begin{equation}
        k = -\epsilon_\lambda\epsilon_v \Gamma_\lambda^2\xi_0^2H_0^2 \left\{ a +B\cos\left[2(s_x-s_u)\right] \right\},
\label{eq:metric_k_constant}
\end{equation}
which is independent of $s$. Once equation~\eqref{eq:metric_k_constraint} is imposed, the other components in equation~\eqref{eq:metric_Einstein_components_required} reduce precisely to equations~\eqref{eq:xprime_final}--\eqref{eq:uprime_final}; hence they generate no additional condition.

On a globally regular branch, equation~\eqref{eq:v_fixed_sign_global_condition} implies $\epsilon_v=-\operatorname{sgn}a$, while the quantity in braces in equation~\eqref{eq:metric_k_constant} has the sign of $a$. It follows that
\begin{equation}
        \operatorname{sgn}k = \epsilon_\lambda =      \operatorname{sgn}(\lambda^2-1).
\label{eq:metric_k_sign}
\end{equation}
The curvature is therefore nonzero on every regular fixed-sign orbit. With the usual normalization $k=\pm1$, the LRS III/B2 representative has $k=-1$, whereas the LRS I/B1 representative has $k=+1$. The constants $H_0$ and $qF_0$ are then fixed by equations~\eqref{eq:metric_k_constant} and \eqref{eq:metric_twist_normalization}, up to coordinate rescalings.

\paragraph{The B1 branch $|\lambda|>1$.} The regular positive-energy B1 branch can be identified more explicitly with the charged Taub--NUT (Brill) geometry \cite{Brill1964,Clement2016}. For $|\lambda|>1$ and $k=1$, define
\begin{equation}
n:=H_0,\qquad \Delta_{\mathrm N}:=s_x-s_u,\qquad c_{\mathrm N}:=a+B\cos(2\Delta_{\mathrm N}).
\end{equation}
The regularity condition $|B|<|a|$ ensures that $c_{\mathrm N}\ne0$ and that it has the sign of $a$. On a local angular patch, introduce
\begin{equation}
 r_{\mathrm N}:=n\tan(s-s_x),\qquad \zeta=2\tan(\theta/2)e^{i\varphi},\qquad T_-:=\frac{n}{q}z-2n\varphi.
\end{equation}
Then 
\begin{equation}
d\Sigma_1^2=d\theta^2+\sin^2\theta\,d\varphi^2 \qquad \omega_1=-2(1-\cos\theta)\,d\varphi\,.
\end{equation} 
Equations~\eqref{eq:metric_twist_normalization} and~\eqref{eq:metric_k_constant} then give
\begin{equation}
 \frac{q^2F^2}{n^2}=-\frac{v}{c_{\mathrm N}} =:f_{\mathrm N},\qquad  H^2=r_{\mathrm N}^2+n^2,\qquad \left(\frac{dr_{\mathrm N}}{dw}\right)^2=f_{\mathrm N}.
\end{equation}
Consequently, the metric takes the form
\begin{equation}
 d\ell^2={} f_{\mathrm N}(r_{\mathrm N}) (dT_-+2n\cos\theta\,d\varphi)^2 -\frac{dr_{\mathrm N}^2}{f_{\mathrm N}(r_{\mathrm N})} -(r_{\mathrm N}^2+n^2)(d\theta^2+\sin^2\theta\,d\varphi^2),
 \end{equation}
where the auxiliary functions are
\begin{equation}
 f_{\mathrm N}(r_{\mathrm N})= \frac{r_{\mathrm N}^2-2M_{\mathrm N}r_{\mathrm N}-n^2 +\mathcal{Q}_{\mathrm N}^{\,2}}{r_{\mathrm N}^2+n^2},\qquad
 M_{\mathrm N}=-\frac{Bn\sin(2\Delta_{\mathrm N})}{c_{\mathrm N}}, \qquad \mathcal{Q}_{\mathrm N}^{\,2}=\frac{2an^2}{c_{\mathrm N}}.
 \end{equation}
Moreover,
\begin{equation}
 \mathcal{Q}_{\mathrm N}^{\,2}-M_{\mathrm N}^2-n^2 =\frac{n^2(a^2-B^2)}{c_{\mathrm N}^2}>0,
\end{equation}
so $f_{\mathrm N}$ has no real zero. This identifies the regular B1 sector locally with the horizonless charged Taub--NUT branch. In the electrovacuum (Brill) interpretation of the metric, $\mathcal{Q}_{\mathrm N}$ is the electric flux through the throat $r_{\mathrm N}=0$, which carries no charged source (``charge without charge'' \cite{MisnerWheeler1957}). In the present construction, $\mathcal{Q}_{\mathrm N}^{\,2}$ labels the geometry and the stress tensor is supplied by the massless Dirac field through the polar relations~\eqref{eq:mu_massless}--\eqref{eq:Q_massless}. The Ricci tensor~\eqref{eq:Ricci_exact_spinor_branch} satisfies the algebraic Rainich conditions $R=0$, $R_a{}^cR_c{}^b=\frac14\delta_a{}^bR_{cd}R^{cd}$ and, for $\mu>0$, $R_{ab}u^au^b\geq0$, which identify the algebraic form of a non-null electromagnetic Ricci tensor \cite{MisnerWheeler1957}. No Maxwell field is introduced here: the comparison is an isometry and stress-tensor equivalence, while the global existence of a spin structure and of a single-valued spinor on a chosen Taub--NUT quotient remains to be checked separately. The result is therefore a massless Dirac source for a known metric, together with the explicit spinorial variables $(\rho,\beta)$ on it, rather than a new geometry.

\paragraph{Non-uniqueness of the spinorial source.} The Brill parameters read
\begin{equation}
        \frac{\mathcal Q_{\mathrm N}^{\,2}}{n^2}=\frac{2a}{a+B\cos(2\Delta_{\mathrm N})}, \qquad \frac{M_{\mathrm N}}{n}=-\frac{b\sin(2a\Delta_{\mathrm N})}{a+B\cos(2\Delta_{\mathrm N})},
\label{eq:brill_parameters_b}
\end{equation}
and do not depend on $\lambda$. For fixed $(n,M_{\mathrm N},\mathcal Q_{\mathrm N})$, every admissible $|\lambda|>1$ therefore yields a regular spinor polar variable whose reconstructed stress tensor reproduces the same metric. Thus, for fixed metric parameters, each admissible $|\lambda|>1$ yields a distinct set of polar variables. These configurations differ in boost rapidity and in the product $\rho\xi=-4\mu/\lambda$. The one-parameter non-uniqueness is established here at the level of the reconstructed polar variables.

\paragraph{Zero-crossing orbits and Killing horizons.} The algebraic relations above also apply on fixed-sign subintervals of orbits with $\mathcal I\geq0$, provided $c_{\mathrm N}\neq0$ and the sheet-curvature normalization is $k=+1$. 
\begin{equation}
        M_{\mathrm N}^2+n^2-\mathcal Q_{\mathrm N}^{\,2}=\frac{n^2\left(B^2-a^2\right)}{c_{\mathrm N}^2}=\frac{n^2\,\mathcal I}{c_{\mathrm N}^2}\geq0,
\label{eq:horizon_condition}
\end{equation}
so $f_{\mathrm N}$ has the real zeros $r_\pm=M_{\mathrm N}\pm\sqrt{M_{\mathrm N}^2+n^2-\mathcal Q_{\mathrm N}^{\,2}}$. These are the Killing horizons of the charged Taub--NUT metric, and they are degenerate (extremal) for $\mathcal I=0$. Since $f_{\mathrm N}=-v/c_{\mathrm N}$, the horizons occur at interior zeros of $v$; there $\xi\to0$ while $\rho\propto|v|^{-1/2}$ diverges. The metric admits an analytic continuation through the horizons, with a nonstationary region between distinct roots when $\mathcal I>0$. Thus, within this nondegenerate $k=+1$ family, $\mathcal I<0$, $\mathcal I=0$ and $\mathcal I>0$ describe horizonless, extremal and non-extremal geometries, respectively. The full-interval regular-spinor condition selects the horizonless class, although regular spinor patches exist away from the zeros on the other orbits.

\paragraph{The B2 branch with $|\lambda|<1$.} The same construction applies to the B2 representative. For $k=-1$ set 
\begin{equation}
\zeta=2\tanh(\theta/2)e^{i\varphi},
\end{equation} 
so that 
\begin{equation}
d\Sigma_{-1}^2=d\theta^2+\sinh^2\theta\,d\varphi^2 \qquad \omega_{-1}=-2(\cosh\theta-1)\,d\varphi\,.
\end{equation}
With $n$, $\Delta_{\mathrm N}$, $c_{\mathrm N}$ and $r_{\mathrm N}$ as above, define $T_+:=(n/q)z+2n\varphi$ on this angular patch. The metric then becomes
\begin{equation}
        d\ell^2= f_{\mathrm N}(r_{\mathrm N})\left(dT_+-2n\cosh\theta\,d\varphi\right)^2-\frac{dr_{\mathrm N}^2}{f_{\mathrm N}(r_{\mathrm N})}-\left(r_{\mathrm N}^2+n^2\right)\left(d\theta^2+\sinh^2\theta\,d\varphi^2\right),
\label{eq:hyperbolic_nut_metric}
\end{equation}
with
\begin{equation}
\begin{split}
f_{\mathrm N}=\frac{v}{c_{\mathrm N}}=\frac{-\left(r_{\mathrm N}^2-n^2\right)-2M_{\mathrm N}r_{\mathrm N}+\mathcal Q_{\mathrm N}^{\,2}}{r_{\mathrm N}^2+n^2}<0,\\ M_{\mathrm N}=\frac{Bn\sin(2\Delta_{\mathrm N})}{c_{\mathrm N}},\qquad \mathcal Q_{\mathrm N}^{\,2}=-\frac{2an^2}{c_{\mathrm N}}<0.
\end{split}
\label{eq:hyperbolic_nut_parameters}
\end{equation}
This is the hyperbolic ($k=-1$) member of the charged NUT family \cite{PlebanskiDemianski1976} with a formal negative squared-charge parameter, corresponding algebraically to the opposite sign of the Maxwell stress tensor, in agreement with $\mu<0$. Here
\begin{equation}
M_{\mathrm N}^2+n^2+\mathcal Q_{\mathrm N}^{\,2}=n^2(B^2-a^2)/c_{\mathrm N}^2<0
\end{equation} 
on the regular branch, so $f_{\mathrm N}$ has no zero and $r_{\mathrm N}$ is a global time coordinate on the chart. The hypersurfaces $r_{\mathrm N}=\mathrm{const}$ are circle (or line) bundles over the hyperbolic plane, homogeneous under a four-dimensional isometry group with three-dimensional subgroup of Bianchi type~VIII; the B2 branch is therefore an LRS Bianchi~VIII cosmology with a phantom Brill-type source. For the elementary orbit $B=0$, one has $M_{\mathrm N}=0$, $\mathcal Q_{\mathrm N}^{\,2}=-2n^2$ and $f_{\mathrm N}=-1$.

\subsection{An elementary representative}
\label{subsec:elementary_metric_representative}

The invariant orbit $\mathcal L_*$ in equation~\eqref{eq:invariant_reference_orbit}, which lifts the critical point $P_*$ of the two-dimensional subsystem, corresponds to $B=0$ and gives a particularly simple member of the reconstructed family:
\begin{equation}
        v=-a, \qquad y=0, \qquad  r=-\frac{1}{a}.
\label{eq:elementary_metric_orbit}
\end{equation}
Define
\begin{equation}
        C:=\Gamma_\lambda|\xi_0| \sqrt{|a|}, \qquad      \epsilon_0:=\operatorname{sgn}\xi_0 =-\operatorname{sgn}a.
\label{eq:elementary_metric_C}
\end{equation}
After setting $w_0=0$, normalizing $k=\epsilon_\lambda$, and rescaling the fibre coordinate, equation~\eqref{eq:canonical_reconstructed_metric} becomes
\begin{equation}
        d\ell^2={} -\epsilon_\lambda\,dw^2 +\epsilon_\lambda        \left[dz+\frac{\epsilon_0}{C} e^{\Psi_{\epsilon_\lambda}}    \operatorname{Im}\left(\zeta\,d\bar\zeta\right)\right]^2-\left(w^2+C^{-2}\right) e^{2\Psi_{\epsilon_\lambda}}      d\zeta\,d\bar\zeta.
\label{eq:elementary_metric_explicit}
\end{equation}
The normalized parameter, polar variables, and independent connection variables are elementary functions of $w$:
\begin{align}
        s(w) &=s_x+\epsilon_0\arctan(Cw),\label{eq:elementary_metric_s_w}\\
        \xi(w) &= \frac{\xi_0\sqrt{|a|}}{1+C^2w^2}, &        \rho(w) &= \frac{|\xi_0|}{\sqrt{|a|}} \frac{1}{1+C^2w^2},\label{eq:elementary_metric_polar_variables}\\
        X(w) &=2\epsilon_0Cw\,\xi(w), & Y(w)&=0,\label{eq:elementary_metric_connection_variables}\\
        \beta(w) &=\beta_0-s(w).
\end{align}
The remaining connection scalars obey $\Omega=\lambda\xi$, $\phi=\lambda X$, and $\mathcal A=0$, while $\Theta$ and $\Sigma$ follow from equation~\eqref{eq:XY_inverse}. The resulting source is
\begin{equation}
        \mu(w) = \frac{2\epsilon_\lambda C^2}{(1+C^2w^2)^2},       \qquad p=\frac{\mu}{3}, \qquad \Pi=-\frac{4\mu}{3},       \qquad Q=0,
\label{eq:elementary_metric_source}
\end{equation}
and direct evaluation gives $R=0$ and $R_{ab}=\mu(P_{ab}-N_{ab})$, in agreement with equation~\eqref{eq:Ricci_exact_spinor_branch}. The orbit is not an equilibrium of the spinor-current dynamical system: $x$ still evolves monotonically along $\mathcal L_*$. The associated metric is an elementary, nonstationary LRS III representative when $|\lambda|<1$, and an elementary stationary LRS I representative when $|\lambda|>1$.

For the positive-energy realization of this elementary orbit, the parameters introduced above reduce to
\[
 n=C^{-1},\qquad M_{\mathrm N}=0,\qquad \mathcal{Q}_{\mathrm N}^{\,2}=2n^2,\qquad f_{\mathrm N}=1.
\]
This is the massless ($M_{\mathrm N}=0$) member with $f_{\mathrm N}\equiv1$, singled out in Ref.~\cite{Clement2016} as the simplest NUT wormhole. Although the metric is stationary, the spinor-current variables $\xi$, $\rho$ and $X$ depend on $w$, because the current moves with constant rapidity relative to the Killing direction $\partial_z$.

The latter observation also clarifies the meaning of the absence of a regular equilibrium in the spinor-current variables. It rules out identifying the spinor-current congruence itself with a rigid, shear-free and expansion-free G\"odel-type flow \cite{Godel1949,Stephani2003}, since $X=Y=0$ would give $\dot X=2\xi^2\neq0$. It does not rule out stationarity of the spacetime in a different congruence. In the $|\lambda|>1$ sector the canonical boosted frame is LRS I, $\widetilde\Theta=\widetilde\Sigma=0$, and the metric coefficients in equation~\eqref{eq:canonical_reconstructed_metric} are independent of the timelike fibre coordinate $z$.

The positive-energy B1 realization has a further global qualification. On the standard $k=+1$ angular chart, write $\zeta=R_\zeta e^{i\varphi}$. A circle at fixed $w$, $z$, and
$R_\zeta$ has squared tangent norm
\[
 g_{\varphi\varphi} =\frac{R_\zeta^2}{(1+R_\zeta^2/4)^2}   \left[q^2F(w)^2R_\zeta^2-H(w)^2\right].
\]
Thus, whenever the coordinate domain contains the complete angular circle with $R_\zeta>H/(|q|F)$, that circle is closed and timelike. For the elementary representative, this condition becomes $R_\zeta^2>1+C^2w^2$. No identification of the timelike fibre coordinate $z$ is needed for these curves to close. A smaller local domain need not contain such a complete circle. Positive energy density therefore does not by itself establish a globally causal B1 realization.

This global qualification is familiar from the Taub--NUT geometry. Bonnor's interpretation retains a nonperiodic time coordinate
and interprets the singular half-axis as a semi-infinite massless source of angular momentum \cite{Bonnor1969}. The charged case and its causal structure are discussed in Ref.~\cite{Clement2016}, which keeps a nonperiodic time, finds the Misner string transparent to geodesic motion, and identifies parameter ranges free of closed timelike and null geodesics. In particular, the angularly dependent shift from $z$ to $T$ used above is not single-valued around a complete azimuthal circle. The local charged Taub--NUT identification therefore leaves the choice of global time and fiber identifications open.

The reconstruction above is local. The original spinor-current frame is recovered by applying the inverse constant boost in equation~\eqref{boost}, but the coordinate ranges, spatial topology, fibre identifications, and global spin structure are not fixed by the covariant scalars. In the LRS III sector $w$ is a natural time coordinate and covers $\mathbb R$ on the regular branch, whereas in the LRS I sector the fibre is timelike and its possible global identifications require particular care. Therefore the explicit metric makes the questions of geodesic completeness and global causality well posed, but does not by itself prove geodesic completeness, the existence of Cauchy surfaces, or global hyperbolicity.

\section{Conclusion}
\label{sec:conclusion}

We have analyzed a massless Dirac field, in the polar (spinorial-fluid) formulation, on LRS spacetimes, with the spinor current and spin direction aligned with the LRS frame. When the effective fluid is perfect, LRS~I admits no solution with $\rho>0$, and LRS~II and III admit only stealth configurations ($T_{ab}=0$ with nonvanishing polar variables). In LRS~III, anisotropic stress and heat flux are excluded as well. In non-perfect LRS~I the heat flux vanishes and the equations close into the free first-order system~\eqref{eq:LRSI_nonperfect_system}, so local solutions with anisotropic stress, including solutions with $\mu>0$, exist. The non-perfect LRS~II sector has not been classified.

When both the vorticity and the sheet twist of the current are nonzero, their ratio $\lambda$ is constant and every LRS scalar depends on a single variable. Integrability of the chiral angle then requires $\Delta=0$ and $\lambda^2\neq1$. Hence $Q=0$ and $\mu+p+\Pi=0$: the null energy condition is saturated along the two radial null directions, and the Ricci tensor takes the form $R_{ab}=\mu(P_{ab}-N_{ab})$, with eigenvalues $\{\mu,\mu,-\mu,-\mu\}$. The reduced system for $X$, $Y$, $\rho$ and $\xi$ has no equilibrium point.

After normalization by the sheet twist, the system reduces to a Riccati equation and a shifted harmonic oscillator, and we integrate it in closed form. On fixed-sign orbits ($|B|<|a|$), the connection scalars and the polynomial curvature invariants are bounded and decay at both ends of the range $\tau\in\mathbb R$ of the proper time of the current. The energy density satisfies $\operatorname{sgn}\mu=\operatorname{sgn}(\lambda^2-1)$. The compactified phase space has no isotropic (FLRW) limit; in the LRS~III range its nondegenerate critical sets correspond to Milne-type and LRS Kasner asymptotic regimes. Such persistence of anisotropy ($\widetilde\Sigma/\widetilde\Theta\to1/3$ at late times) does not conflict with Wald's cosmic no-hair theorem \cite{Wald1983} nor the analysis of Collins and Hawking \cite{CollinsHawking1973}, as in our case we have $\Lambda=0$ and, since $T_{ab}t^at^b$ has the sign of $\mu<0$ for every timelike congruence $t^a$, the dominant and strong energy conditions fail on this branch.

A constant boost maps the current to a stationary LRS~I congruence for $|\lambda|>1$ and to an irrotational, geodesic LRS~III congruence for $|\lambda|<1$. In these frames, the metric is reconstructed in the Goode--Wainwright--Van den Bergh form B1 with $k>0$ or B2 with $k<0$, respectively. For $|\lambda|>1$ it is locally isometric to the horizonless charged Taub--NUT (Brill) metric, with $\mathcal Q_{\mathrm N}^{\,2}>M_{\mathrm N}^2+n^2$; the massless Dirac field provides its stress--energy tensor, in the polar representation, without the introduction of a Maxwell field. The invariant orbit $\mathcal L_*$ ($B=0$) gives the member with $M_{\mathrm N}=0$ and $\mathcal Q_{\mathrm N}^{\,2}=2n^2$. For $|\lambda|<1$ the energy density is negative and the null energy condition is violated in every non-radial null direction; the metric is then the hyperbolic ($k=-1$) charged NUT metric with $\mathcal Q_{\mathrm N}^{\,2}<0$, an LRS Bianchi~VIII cosmology. For nondegenerate $k=+1$ continuations with $c_{\mathrm N}\neq0$, the excluded orbits $|B|\geq|a|$ have Killing horizons (extremal for $|B|=|a|$) at interior zeros of $v$, where the polar modulus diverges. The Brill parameters depend only on $B/a$ and $s_x-s_u$, not on $\lambda$; each metric therefore admits a one-parameter family of distinct reconstructed polar configurations.

All metric identifications are local, and bounded curvature on fixed-sign orbits is not claimed to imply geodesic completeness or global hyperbolicity. On the full angular domain the B1 metric contains closed timelike curves, as is familiar for Taub--NUT; a global interpretation requires a choice between Misner's periodic identification of the fiber and Bonnor's string interpretation, together with the existence of a spin structure and of a single-valued spinor on the resulting manifold. For $|\lambda|<1$ the hypersurfaces orthogonal to the boosted congruence define a time function on the region covered by the chart; whether they are Cauchy surfaces has not been established. 

\section*{Acknowledgements}
EB is partially supported by \textit{Conselho Nacional de Desenvolvimento Científico e Tecnológico} (grant N.\ 306305/2026-7) and FAPEMIG (APQ-05207-23). PKSD is supported by a grant from First Rand Bank (SA). This work has been carried out in the framework of activities of the INFN Research Project QGSKY.

\appendix

\section{Specialized LRS system and reduction}
\label{app:lrs_system}

For completeness, this appendix collects the LRS equations used in the main text. They follow from the $1+1+2$ equations in the signature $(+---)$ and the conventions of Ref.~\cite{Vignolo2025}. The sign of $Q$ is fixed by equation~\eqref{eq:T_decomposition}.

The evolution equations are
\begin{align}
\dot\Omega &=-\mathcal A\xi-\frac{2}{3}\Theta\Omega-\Omega\Sigma,\label{eq:app_dot_Omega}\\
\dot{\mathcal H} &=3\mathcal E\xi+\frac{3}{2}\Pi\xi-\Theta\mathcal H -\frac{3}{2}\Sigma\mathcal H-\Omega Q,\label{eq:app_dot_H}\\
\dot\phi &=-\left(\Sigma+\frac{2}{3}\Theta\right) \left(\mathcal A+\frac{1}{2}\phi\right) +2\Omega\xi-Q,\label{eq:app_dot_phi}\\
\dot\xi &=-\left(\frac{1}{2}\Sigma+\frac{1}{3}\Theta\right)\xi -\Omega\left(\frac{1}{2}\phi+\mathcal A\right)+\frac{1}{2}\mathcal H.\label{eq:app_dot_xi}
\end{align}
The propagation equations are
\begin{align}
\hat\Omega &=-\Omega(\mathcal A+\phi), \label{eq:app_hat_Omega}\\
\hat\phi &=-\frac{1}{2}\phi^2+2\xi^2 -\left(\Sigma-\frac{1}{3}\Theta\right) \left(\Sigma+\frac{2}{3}\Theta\right) -\frac{2}{3}\mu-\frac{1}{2}\Pi+\mathcal E, \label{eq:app_hat_phi}\\
\hat\xi &=-\left(\Sigma-\frac{1}{3}\Theta\right)\Omega-\phi\xi, \label{eq:app_hat_xi}\\
\frac{2}{3}\hat\Theta+\hat\Sigma &=-\frac{3}{2}\Sigma\phi+2\Omega\xi-Q, \label{eq:app_hat_Theta_Sigma}\\
\hat{\mathcal E}-\frac{1}{2}\hat\Pi+\frac{1}{3}\hat\mu &=-\frac{3}{2}\phi\left(\mathcal E-\frac{1}{2}\Pi\right)-3\Omega\mathcal H -Q\left(\frac{1}{3}\Theta+\frac{1}{2}\Sigma\right), \label{eq:app_hat_E}\\
\hat{\mathcal H} &=-\frac{3}{2}\phi\mathcal H -\Omega\left(-3\mathcal E+\mu+p-\frac{1}{2}\Pi\right) +Q\xi. \label{eq:app_hat_H}
\end{align}
The mixed evolution--propagation equations and conservation laws are
\begin{align}
\dot\Theta+\hat{\mathcal A} &=-\mathcal A\phi+\mathcal A^2-\frac{1}{3}\Theta^2 -\frac{3}{2}\Sigma^2+2\Omega^2-\frac{1}{2}(\mu+3p),\label{eq:app_Raychaudhuri}\\
\dot\mu-\hat Q &=-(\mu+p)\Theta+Q\phi-2\mathcal A Q+\frac{3}{2}\Sigma\Pi, \label{eq:app_energy_conservation}\\
\dot Q-\hat\Pi-\hat p &=\frac{3}{2}\Pi\phi-\mathcal A(\mu+p+\Pi) -\frac{4}{3}Q\Theta+Q\Sigma, \label{eq:app_momentum_conservation}\\
\dot\Sigma-\frac{2}{3}\hat{\mathcal A} &=-\frac{1}{3}\mathcal A\phi+\frac{1}{2}\Sigma^2 +\frac{2}{3}\Omega^2-\frac{2}{3}\mathcal A^2 -\frac{2}{3}\Theta\Sigma-\mathcal E-\frac{1}{2}\Pi, \label{eq:app_dot_Sigma}\\
\dot{\mathcal E}-\frac{1}{2}\dot\Pi+\frac{1}{3}\hat Q &=-\mathcal E\Theta-\frac{3}{2}\mathcal E\Sigma-3\mathcal H\xi +\frac{2}{3}\mathcal A Q+\frac{1}{6}Q\phi \nonumber\\
&\quad -\frac{1}{2}(\mu+p)\Sigma +\frac{1}{6}\Theta\Pi-\frac{1}{4}\Sigma\Pi. \label{eq:app_dot_E}
\end{align}
The algebraic magnetic-Weyl constraint is
\begin{equation}
        \mathcal H=2\mathcal A\Omega+\Omega\phi-3\xi\Sigma.
\label{eq:app_H_constraint}
\end{equation}

Every LRS scalar satisfies
\begin{align}
        \Omega\dot f&=\xi\hat f, \label{eq:app_scalar_relation}\\
        \widehat{\dot f}-\dot{\hat f} &=\mathcal A\dot f -\left(\Sigma-\frac{1}{3}\Theta\right)\hat f. \label{eq:app_scalar_commutator}
\end{align}
The geometrical and matter consistency conditions needed below are
\begin{align}
        \phi\xi &=\left(\Sigma+\frac{2}{3}\Theta\right)\Omega, \label{eq:app_geometric_constraint}\\
        (\mu+p+\Pi)\xi\Omega &=Q(\Omega^2+\xi^2). \label{eq:app_matter_constraint}
\end{align}
For constant $\lambda$ and equation~\eqref{eq:generic_geometric_relations_setup}, the commutator gives the derived, rather than independent, relation
\begin{equation}
        \hat{\hat f}-\lambda^2\ddot f=0.
\label{eq:characteristic_wave_equation}
\end{equation}

In the massless polar-spinor sector, the substitutions are
\begin{align}
\mu&=\frac{\rho}{2}\left(-\frac{1}{2}\hat\beta-\Omega\right), & p&=-\frac{\rho}{12}(\hat\beta+2\Omega), \label{eq:app_massless_mu}\\
\Pi&=-\frac{\rho}{6}(\hat\beta-\Omega), & Q&=-\frac{\rho}{4}(\dot\beta+\xi), \label{eq:app_massless_matter}\\
\dot\rho&=-\Theta\rho, & \hat\rho&=(\mathcal A-\phi)\rho. \label{eq:app_massless_density}
\end{align}

We finally summarize how this overdetermined-looking system reduces to the four equations in Sec.~\ref{sec:generic_branch}.
\begin{enumerate}
\item Equation~\eqref{eq:app_geometric_constraint} gives $\phi=\lambda X$. Applying equation~\eqref{eq:app_scalar_relation} to $\ln\rho$ and using equation~\eqref{eq:app_massless_density} then gives $\mathcal A=\lambda Y$.
\item Equations~\eqref{eq:app_dot_Omega}, \eqref{eq:app_dot_xi}, and \eqref{eq:app_H_constraint} yield equation~\eqref{eq:lambda_constancy_derivation}; hence $\lambda$ is constant. The same magnetic-Weyl constraint gives equation~\eqref{eq:H_Delta}.
\item Equations~\eqref{eq:app_matter_constraint}, \eqref{eq:app_hat_phi}, and \eqref{eq:app_massless_matter} give the algebraic reconstruction \eqref{eq:Pi_Delta}--\eqref{eq:E_Delta}. Applying equation~\eqref{eq:app_scalar_relation} to $\beta$ then gives equation~\eqref{eq:branch_constraint}.
\item On the regular branch $\Delta=0$, equations~\eqref{eq:app_dot_phi}, \eqref{eq:app_Raychaudhuri}, \eqref{eq:app_massless_density}, and \eqref{eq:app_dot_xi} yield equations~\eqref{eq:final_X}--\eqref{eq:final_xi}. Equation~\eqref{eq:app_momentum_conservation} propagates $\Pi=\lambda\rho\xi/3$.
\item After these substitutions, equations~\eqref{eq:app_dot_H}, \eqref{eq:app_hat_Omega}--\eqref{eq:app_hat_H}, \eqref{eq:app_energy_conservation}, \eqref{eq:app_dot_Sigma}, and \eqref{eq:app_dot_E} reduce to identities. 
\end{enumerate}

\section{Reconstruction of the asymptotic LRS III power laws}
\label{app:critical_point_reconstruction}

This appendix derives the geometrical identifications of the nondegenerate critical sets at infinity directly from the twist-normalized variables. The construction applies to the boosted LRS III frame, $|\lambda|<1$, where the field equations are evolutionary. In the LRS I regime, the same projective sets instead describe spatial propagation limits, so no temporal power-law interpretation is implied.

Let $t$ be proper time along $\widetilde u^a$ and set $\gamma=\sqrt{1-\lambda^2}$. From equations~\eqref{eq:XY_inverse}, \eqref{potentials}, and
\eqref{eq:normalized_variables_final},
\begin{equation}
\widetilde\Theta=\gamma\xi(x-y), \qquad \widetilde\Sigma=\frac{\gamma\xi}{3}(x+2y).
\label{eq:app_projective_theta_sigma}
\end{equation}
The eigenvalues of the expansion tensor along the preferred spatial direction and the two equivalent sheet directions are
\begin{equation}
H_\parallel =\frac{\widetilde\Theta}{3}-\widetilde\Sigma =-\widetilde Y, \qquad H_\perp =\frac{\widetilde\Theta}{3} +\frac{\widetilde\Sigma}{2} =\frac{\widetilde X}{2}.
\label{eq:app_directional_hubble_rates}
\end{equation}
Consequently,
\begin{equation}
    \frac{H_\parallel}{\widetilde\Theta} =-\frac{y}{x-y}, \qquad \frac{H_\perp}{\widetilde\Theta} =\frac{x}{2(x-y)}.
\label{eq:app_directional_rate_ratios}
\end{equation}

Approach a nondegenerate point of the Poincar\'e sphere by writing $\boldsymbol q=(x,y,r)=\mathcal{R}\boldsymbol z+o(\mathcal{R})$, with $\mathcal{R}\to\infty$ and $z_x-z_y\neq0$. Equations~\eqref{eq:app_projective_theta_sigma} and \eqref{eq:app_directional_rate_ratios} give
\begin{align}
        \frac{\widetilde\Sigma}{\widetilde\Theta} &\longrightarrow \frac{z_x+2z_y}{3(z_x-z_y)}, \label{eq:app_projective_shear_ratio}\\
        \frac{H_\parallel}{\widetilde\Theta} &\longrightarrow -\frac{z_y}{z_x-z_y}, & \frac{H_\perp}{\widetilde\Theta} &\longrightarrow        \frac{z_x}{2(z_x-z_y)}. \label{eq:app_projective_directional_ratios}
\end{align}
The last two ratios sum as $H_\parallel/\widetilde\Theta +2H_\perp/\widetilde\Theta=1$, as required.

It remains to determine the time dependence of $\widetilde\Theta$. From equations~\eqref{eq:matter_Delta_zero} and \eqref{eq:app_projective_theta_sigma}, one obtains
\begin{equation}
        \frac{\mu}{\widetilde\Theta^2} =- \frac{\lambda r}{4(1-\lambda^2)(x-y)^2} =O(\mathcal{R}^{-1}),
\label{eq:app_normalized_mu_decay}
\end{equation}
and the same decay holds for $p/\widetilde\Theta^2$ and $\Pi/\widetilde\Theta^2$. Furthermore,
\begin{equation}
        \frac{\widetilde\xi}{\widetilde\Theta} =\frac{1}{x-y}=O(\mathcal{R}^{-1}).
\label{eq:app_normalized_twist_decay}
\end{equation}
Thus both the source and the sheet twist are subleading at every nondegenerate boundary set.

In the boosted LRS III frame, $\widetilde{\mathcal A}=\widetilde\Omega=0$. The Raychaudhuri equation therefore has the asymptotic form
\begin{equation}
        D_{\widetilde u}\widetilde\Theta =- \left[\frac13 + \frac32\left(\frac{\widetilde\Sigma}{\widetilde\Theta}\right)^2 \right]        \widetilde\Theta^2 +o(\widetilde\Theta^2).
\label{eq:app_asymptotic_raychaudhuri}
\end{equation}
At a fixed projective direction, we define
\begin{equation}
        q_\Sigma :=\frac{z_x+2z_y}{3(z_x-z_y)}, \qquad c:=\frac13+\frac32q_\Sigma^2.
        \label{eq:app_c_definition}
\end{equation}
Integration of equation~\eqref{eq:app_asymptotic_raychaudhuri} gives, to leading order,
\begin{equation}
        \widetilde\Theta(t) =\frac{1}{c(t-t_0)}.
\label{eq:app_theta_power_law}
\end{equation}

Let $L_\parallel$ and $L_\perp$ denote local directional scale factors, so that $D_{\widetilde u}\ln L_\parallel=H_\parallel$ and $D_{\widetilde u}\ln L_\perp=H_\perp$. Combining equations~\eqref{eq:app_projective_directional_ratios} and \eqref{eq:app_theta_power_law} yields
\begin{equation}
    L_\parallel\propto|t-t_0|^{\kappa_\parallel}, \qquad L_\perp\propto|t-t_0|^{\kappa_\perp},
    \end{equation}
with
\begin{equation}
    \kappa_\parallel =-\frac{z_y}{c(z_x-z_y)},\qquad \kappa_\perp =\frac{z_x}{2c(z_x-z_y)}.
\label{eq:app_projective_exponents}
\end{equation}

The evaluation on each nondegenerate critical set is immediate:
\begin{align}
    M_\pm:\quad &(z_y=0), &q_\Sigma&=\frac13,  &c&=\frac12,  &(\kappa_\parallel,\kappa_\perp,\kappa_\perp)&=(0,1,1), \nonumber\\
    \mathcal C_-:\quad &(z_x=0,\ z_y\neq0), &q_\Sigma&=-\frac23, &c&=1, &(\kappa_\parallel,\kappa_\perp,\kappa_\perp)&=(1,0,0), \nonumber\\
    \mathcal C_+:\quad &(z_y=z_x/4,\ z_x\neq0), &q_\Sigma&=\frac23, &c&=1, &(\kappa_\parallel,\kappa_\perp,\kappa_\perp) &=\left(-\frac13,\frac23,\frac23\right). \label{eq:app_critical_exponents}
\end{align}
The first triple is a flat Milne-type representative, the second is the flat LRS Kasner or Taub representative, and the third is the non-flat LRS Kasner solution. At $D_\pm$, $z_x=z_y=0$, so neither the expansion scale nor the directional ratios are defined at leading projective order. This is precisely the degeneracy associated with the double-zero boundary of the nonregular spinor sector.

\section{Phase-aligned elliptic time quadrature}
\label{app:elliptic_quadrature}

Only the relative phase $s_x-s_u$ in the exact solution is essential, because one phase can always be absorbed into a shift of $s$. For the phase-aligned subfamily we choose $s_x=s_u=0$ and define
\begin{equation}
        \sigma:=\tan s, \qquad \alpha_{\rm e}:=B-a, \qquad \chi_{\rm e}:=B+a. 
    \label{eq:AC_t_def_time_map}
\end{equation}
The time quadrature becomes
\begin{equation}
        \tau-\tau_0 = \frac{1}{\xi_0} \int^{\sigma} \frac{\sqrt{1+z^2}}{\sqrt{|\alpha_{\rm e}-\chi_{\rm e}z^2|}}\,dz.
\label{eq:tau_t_elliptic_integral}
\end{equation}
On a branch where
\begin{equation}
        \varepsilon:=\operatorname{sgn}(\alpha_{\rm e}-\chi_{\rm e}\sigma^2)
\end{equation}
is constant, an antiderivative is
\begin{equation}
        \tau-\tau_0 = \frac{1}{\xi_0\sqrt{\varepsilon\chi_{\rm e}}} E\!\left( \arcsin\!\left[ \sqrt{\frac{\chi_{\rm e}}{\alpha_{\rm e}}}\tan s \right] \,\middle|\,  -\frac{\alpha_{\rm e}}{\chi_{\rm e}} \right),
\label{eq:tau_incomplete_elliptic_E}
\end{equation}
where
\begin{equation}
        E(\varphi|m):= \int_0^\varphi \sqrt{1-m\sin^2\vartheta}\,d\vartheta
\label{eq:incomplete_elliptic_E_definition}
\end{equation}
is the incomplete elliptic integral of the second kind. Equation~\eqref{eq:tau_incomplete_elliptic_E} is understood by analytic continuation when one of the displayed square roots is imaginary; on a fixed real branch its derivative is the real integrand in equation~\eqref{eq:tau_t_elliptic_integral}.

On the aligned branch with $v>0$ throughout, $\alpha_{\rm e}>0$ and $\chi_{\rm e}<0$. Setting $d_{\rm e}:=-\chi_{\rm e}>0$ gives the manifestly real form
\begin{equation}
        \tau-\tau_0 = \frac{1}{\xi_0\sqrt{d_{\rm e}}} \mathcal E_{\rm h}\!\left(\operatorname{arsinh}\!\left[\sqrt{\frac{d_{\rm e}}{\alpha_{\rm e}}}\tan s \right] \,\middle|\, \frac{\alpha_{\rm e}}{d_{\rm e}}\right),
\label{eq:tau_hyperbolic_elliptic_E}
\end{equation}
where
\begin{equation}
        \mathcal E_{\rm h}(w|m):= \int_0^w\sqrt{1+m\sinh^2 z}\,dz =-iE(iw|m).
\label{eq:hyperbolic_elliptic_E_definition}
\end{equation}
The inverse map is
\begin{equation}
        s(\tau) = \arctan\!\left\{\sqrt{\frac{\alpha_{\rm e}}{d_{\rm e}}}\sinh\!\left[ \mathcal E_{\rm h}^{-1}\!\left(\xi_0\sqrt{d_{\rm e}}\,(\tau-\tau_0) \,\middle|\, \frac{\alpha_{\rm e}}{d_{\rm e}} \right) \right] \right\}.
\label{eq:s_inverse_hyperbolic_elliptic_E}
\end{equation}
Although $\mathcal E_{\rm h}^{-1}$ is not elementary, it is single-valued on the monotonic branch selected above.

For the aligned branch with $v<0$ throughout, $\alpha_e<0$ and $\chi_e>0$. Setting $A_e:=-\alpha_e>0$ and $D_e:=\chi_e>0$ gives the corresponding manifestly real representation by replacing $(\alpha_e,d_e)$ with $(A_e,D_e)$ in equations~\eqref{eq:tau_hyperbolic_elliptic_E}--\eqref{eq:s_inverse_hyperbolic_elliptic_E}. Finally, if $\chi_{\rm e}=0$, the time map is elementary:
\begin{equation}
        \tau-\tau_0 = \frac{1}{2\xi_0\sqrt{|\alpha_{\rm e}|}} \left[\tan s\sqrt{1+\tan^2s} +\operatorname{arsinh}(\tan s) \right],
\label{eq:tau_special_C_zero}
\end{equation}
and $|\tau|\propto\delta^{-2}$ at the endpoints. By contrast, if $\alpha_{\rm e}=0$, then $v$ vanishes at $s=0$ and the two half-intervals are dynamically disconnected in the $s$ parametrization. Each half-interval may still cover an infinite range of $\tau$, but $\rho\propto|s|^{-1}$, so this case does not belong to the regular-spinor branch.

\section*{Data availability statement}
No new data were created or analyzed in this study.

\section*{References}

\bibliography{ref}
\end{document}